\documentclass[prl,twocolumn,superscriptaddress]{revtex4-2}
\usepackage{upgreek}
\usepackage{amssymb}
\usepackage{amsfonts}
\usepackage{bbm}
\usepackage{mathrsfs}
\usepackage{graphics,graphicx,epsfig,bm,amsmath,amsthm,amssymb}
\usepackage{bm}
\usepackage{longtable}
\usepackage{multirow}
\usepackage{array}
\usepackage{soul}
\soulregister\cite7
\soulregister\ref7
\usepackage{color}

\usepackage{cases}
\usepackage{booktabs}
\usepackage[usenames,dvipsnames]{xcolor}

\usepackage[none]{hyphenat}
\usepackage{float}
\usepackage[a4paper,colorlinks=true,
linkcolor=blue,citecolor=blue,
pdfauthor={ },
pdftitle={ },
pdfsubject={ },
pdfkeywords={ }]{hyperref}
\usepackage{dcolumn}

\begin{document}
	
	\preprint{APS/123-QED}
	
	\title{New Constraints on Dark Photon Dark Matter with a Millimeter-Wave Dielectric Haloscope}
	
	\author{Guoqing Wei}
	\affiliation{Laboratory of Spin Magnetic Resonance, School of Physical Sciences, Anhui Province Key Laboratory of Scientific Instrument Development and Application, University of Science and Technology of China, Hefei 230026, China}
	\affiliation{Hefei National Laboratory, University of Science and Technology of China, Hefei 230088, China}
	
	\author{Diguang Wu}
	\affiliation{Laboratory of Spin Magnetic Resonance, School of Physical Sciences, Anhui Province Key Laboratory of Scientific Instrument Development and Application, University of Science and Technology of China, Hefei 230026, China}
	\affiliation{Hefei National Laboratory, University of Science and Technology of China, Hefei 230088, China}
	
	\author{Runqi Kang}
	\affiliation{Laboratory of Spin Magnetic Resonance, School of Physical Sciences, Anhui Province Key Laboratory of Scientific Instrument Development and Application, University of Science and Technology of China, Hefei 230026, China}
	\affiliation{Hefei National Laboratory, University of Science and Technology of China, Hefei 230088, China}
	
	\author{Qingning Jiang}
	\affiliation{Laboratory of Spin Magnetic Resonance, School of Physical Sciences, Anhui Province Key Laboratory of Scientific Instrument Development and Application, University of Science and Technology of China, Hefei 230026, China}
	
	\author{Man Jiao}
	\email{man.jiao@zju.edu.cn}
	\affiliation{Institute for Advanced Study in Physics and School of Physics, Zhejiang University, Hangzhou 310027, China}
	\affiliation{State Key Laboratory of Ocean Sensing and School of Physics, Zhejiang University, Hangzhou 310058, China}
	\affiliation{Institute of Quantum Sensing, Institute of Fundamental and Transdisciplinary Research and Zhejiang Key Laboratory of R$\&$D and Application of Cutting-edge Scientific Instruments, Zhejiang University, Hangzhou 310058, China}
	
	\author{Xing Rong}%
	\email{xrong@ustc.edu.cn}
	\affiliation{Laboratory of Spin Magnetic Resonance, School of Physical Sciences, Anhui Province Key Laboratory of Scientific Instrument Development and Application, University of Science and Technology of China, Hefei 230026, China}
	\affiliation{Hefei National Laboratory, University of Science and Technology of China, Hefei 230088, China}
	\affiliation{State Key Laboratory of Ocean Sensing and School of Physics, Zhejiang University, Hangzhou 310058, China}
	\affiliation{Institute of Quantum Sensing, Institute of Fundamental and Transdisciplinary Research and Zhejiang Key Laboratory of R$\&$D and Application of Cutting-edge Scientific Instruments, Zhejiang University, Hangzhou 310058, China}
	
	\author{Jiangfeng Du}%
	\affiliation{Hefei National Laboratory, University of Science and Technology of China, Hefei 230088, China}
	\affiliation{State Key Laboratory of Ocean Sensing and School of Physics, Zhejiang University, Hangzhou 310058, China}

	
	\begin{abstract}
		Dark matter remains one of the most profound and unresolved mysteries in modern physics. To unravel its nature, numerous haloscope experiments have been implemented across various mass ranges. However, very few haloscope experiments conducted within millimeter-wave frequency range, which is in the favored mass region for well-motivated dark matter candidates. Here we designed and constructed a millimeter-wave dielectric haloscope featuring a dark matter detector composed of dielectric disks and a mirror. Using this setup, we conducted a search for randomly polarized dark photon dark matter and found no evidence for its existence. Our results established new constraints on the kinetic mixing parameter in the mass range from $387.72$ to $391.03\,\rm{\upmu eV}$, improving the existing limits by two orders of magnitude. With future enhancements, our system has the potential to explore new parameter space for dark photon as well as axion dark matter within the millimeter-wave frequency range.
	\end{abstract}
	
	\maketitle
	An abundance of astrophysical observation results \cite{rubin_rotation_1970,rubin_rotational_1982,begeman_extended_1991,tyson_detailed_1998,taylor_gravitational_1998,clowe_direct_2006} indicate that the majority of the mass of the universe exists in some unknown form, which is called dark matter  \cite{trimble_existence_1987,feng_dark_2010,bertone_history_2018}. Although the nature of dark matter remains elusive, numerous hypotheses have been proposed to explain its composition. Among these, dark photon, a hypothetical spin-1 gauge boson, has garnered significant attention in recent years \cite{battaglieri_us_2017,billard_direct_2022,caputo_dark_2021,cyncynates_experimental_2025}. As an ultralight dark matter candidate, dark photons are theoretically well-motivated and possess varied production mechanisms \cite{graham_vector_2016,dror_parametric_2019,agrawal_relic_2020,ahmed_gravitational_2020,co_dark_2019,cyncynates_detectable_2025}. Dark photons are characterized by the kinetic mixing with the ordinary photon as follows \cite{holdom_searching_1986,holdom_two_1986}:
	\begin{equation}
		\mathcal{L} = -\frac{1}{4} F_{\mu\nu}^2 -\frac{1}{4} V_{\mu\nu}^2 - \frac{1}{2}m^2_{A'}A'_{\mu}A'^{\mu} + \frac{1}{2}\chi F_{\mu\nu}V^{\mu\nu},
		\label{eq:one}
	\end{equation}
	where $A_{\mu}$ and $A'_{\mu}$ denote the fields of ordinary photon and dark photons, respectively, $F_{\mu\nu}=\partial_{\mu}A_{\nu}-\partial_{\nu}A_{\mu}$ and $V_{\mu\nu}=\partial_{\mu}A'_{\nu}-\partial_{\nu}A'_{\mu}$ are the corresponding field tensors. Kinetic mixing $\chi$ and dark photon mass $m_{A'}$ construct the parameter space of dark photon. Dark photons could be produced by quantum fluctuations during inflation \cite{graham_vector_2016}. In this mechanism, well-motivated dark photon mass as $m_{A'}\gtrsim10^{-5}\, \rm{eV}$ (i.e., $f_{\rm{DM}}\gtrsim2\,\rm{GHz}$) is yielded from the cosmological constraints on Hubble scale \cite{akrami_planck_2020}. 
	
	Dark photon dark matter can be detected through the kinetic mixing with ordinary photons. This kind of interaction will modify the Maxwell equations by introducing a “dark” current density $\mathbf{J}_{\rm{DM}}$ as \cite{huang_extension_2019,marocco_dark_2021}
	\begin{equation}
		\mathbf{J}_{\rm{DM}} = \chi \epsilon _0 \left(\frac{\partial \mathbf{E_{\rm{DM}}}}{\partial t} - c^2 \nabla \times \mathbf{B_{\rm{DM}}}\right),
		\label{eq:two}
	\end{equation}
	where $\mathbf{E_{\rm{DM}}}=\partial_t \mathbf{A'}$ and $\mathbf{B_{\rm{DM}}}=\nabla \times \mathbf{A'}$ are dark photon electric and magnetic field, respectively, $\epsilon _0$ is the vacuum permittivity and $c$ is the speed of light. The spatial derivative term $\nabla \times \mathbf{B_{\rm{DM}}}$ can be neglected since the de Broglie wavelength of dark photons is much larger than the experimental setup \cite{millar_dielectric_2017_1}. The “dark” current density oscillates at the frequency of $f_{\rm{DM}}\approx m_{A'}c^2/h$. This “dark” current density will induce ordinary photons with the same frequency of $f_{\rm{DM}}$ within the probe, such as a cavity which is the most widely used in haloscope experiments. Many experiments have been implemented to search for dark photons at different mass ranges \cite{chiles_new_2022,fan_one-electron_2022,an_direct_2023,knirck_first_2024,kang_near-quantum-limited_2024,tang_first_2024,madmax_collaboration_first_2025}, yet the millimeter-wave frequency range remains rarely explored. It is challenging to conduct haloscope experiment with cavity at such high frequency, due to the deterioration of quality factor and decrease of effective mode volume with increasing frequency. Many alternative methods aiming high frequency range have been proposed, such as dish antenna \cite{knirck_first_2024}, metal plate \cite{kotaka_search_2023}, dielectric haloscope \cite{caldwell_dielectric_2017,de_miguel_dark_2021} and quantum cyclotron \cite{fan_one-electron_2022,fan_highly_2025}. The dielectric haloscope is a promising method due to the wide search band and high dark-photon-to-photon conversion rate, and careful calibration and analysis are required to estimate its sensitivity. Recently, dielectric haloscopes have been utilized to search for dark photons in the centimeter-wave \cite{madmax_collaboration_first_2025,cervantes_search_2022} and optical frequency ranges \cite{chiles_new_2022,manenti_search_2022}, while the millimeter-wave range remains untouched. There are new challenges in the implementation of dielectric haloscope in the millimeter-wave frequency range, primarily due to the significantly smaller spacing between the dielectric disks compared to centimeter-wave haloscopes.
	
	\begin{figure}[t]
		\includegraphics[width=1\linewidth]{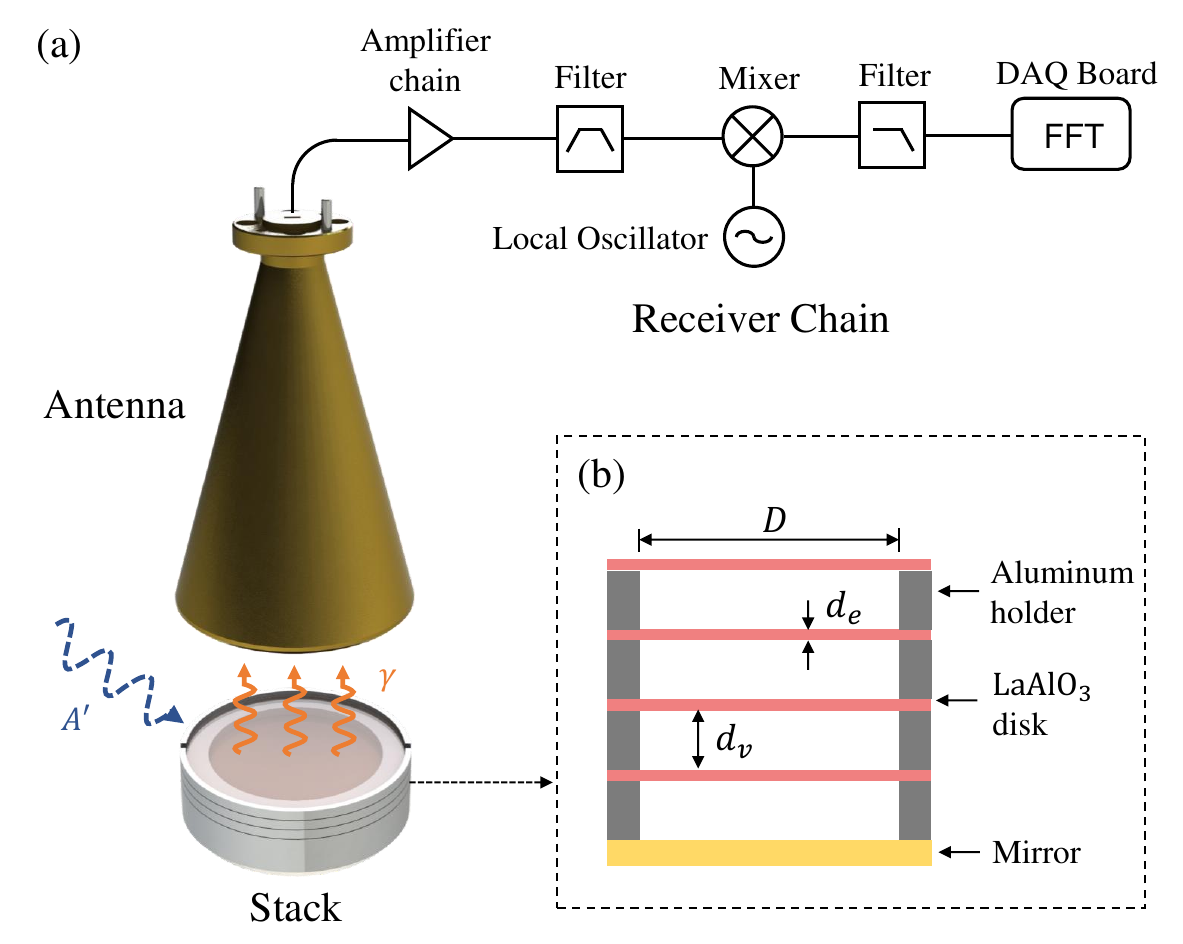}
		\caption{\label{fig:1} Experimental Setup. (a) Simplified diagram of the experiment apparatus for the dark photon search. $A'$ and $\gamma$ stand for the dark photon field and the converted photons, respectively. The emitted photons are collected by an antenna with an aperture of $39.2\,\rm{mm}$. The receiver chain can detect the signal with high efficiency. (b) Diagram of the dielectric stack. The stack is comprised of one gold-coating aluminum mirror and four $\rm{LaAlO_3}$ disks with refractive index $n \approx 5$. The disks and mirror are separated by aluminum holders. The inner diameter of the holder is $D\approx32.0\,\rm{mm}$, while $d_v\approx1.580\, \rm{mm}$ and $d_e\approx0.320\, \rm{mm}$ denote the spacing and thickness of the disks, respectively.}
	\end{figure}
	
	In this paper, we designed and implemented the first dielectric haloscope operating in the millimeter-wave frequency range. To convert dark photons to photons with high efficiency, we constructed a millimeter-wave stack consisting of four dielectric disks and a mirror. A linearly polarized antenna and a millimeter-wave receiver chain were implemented to detect the dark photon signal with low noise. In order to obtain the boost factor of this dielectric haloscope, we developed a method to precisely characterize the stack as well as the antenna. Additionally, to enhance the dark photon search efficiency, we employed a data acquisition (DAQ) board based on the field programmable gate array (FPGA), enabling a duty cycle approaching $100\%$. With the total acquisition time of eight days, our prototype established new constraints on kinetic mixing $\chi$ in the mass range from $387.72$ to $391.03\,\rm{\upmu eV}$, corresponding to the frequency from $93.750$ to $94.550 \,\rm{GHz}$.

	Our haloscope apparatus consisted of three parts: dielectric stack, antenna and receiver chain, as shown in Fig.\,\ref{fig:1}(a). The dielectric stack served as the fundamental component in which dark photons were converted to ordinary photons. As shown in Fig.\,\ref{fig:1}(b), four parallel dielectric disks and a mirror were meticulously arranged with a particular spacing $d_v$, which was determined by the aluminum holders. Driven by the “dark” current density $\mathbf{J_{\rm{DM}}}$, photons were emitted at each interface between the dielectric material and air due to the different refractive indices \cite{millar_dielectric_2017}. We chose a half-wave stack configuration, meaning that photons with center frequency $f_c$ would accumulate a $\pi$ phase shift when passing through each disk or air gap. Thus, the photons converted at each interface would be coherently superposed, resulting in a conversion rate that scaled with the square of dielectric disk number, $N^2$. We positioned a gold-coating mirror on one side of the dielectric stack to reflect photons emitted in that direction, and achieved a fourfold increase in the overall conversion rate on the opposite side. In this dielectric haloscope, the power collected by the antenna can be expressed as \cite{horns_searching_2013,knirck_first_2024}
	\begin{equation}
		P = \chi ^2 c \rho_{\rm{DM}} A \langle\cos^2{\theta}\rangle \lvert \mathcal{B} \rvert^2 ,
		\label{eq:three}
	\end{equation}
	where $\rho_{\rm{DM}} = \epsilon _0 \lvert \mathbf{E_{\rm{DM}}} \rvert^2 /2$ is the local density of dark photon dark matter, and in this paper $\rho_{\rm{DM}}$ is set as $0.45 ~\rm{GeV/cm^3}$ \cite{read_local_2014}. The parameter $A$ denotes the area of the dielectric disk, and $\theta$ is the angle between the polarization vectors of dark photons and experimental receiver system. We assume that dark photons are randomly polarized, yielding an average value of $\langle\cos^2{\theta}\rangle = 1/3$ \cite{ghosh_searching_2021}. The parameter $\lvert \mathcal{B} \rvert ^2$ represents the boost factor of the signal due to the constructive interference, which is dependent on the frequency of dark photons. Assuming an ideal antenna and a perfect half-wave dielectric stack, the boost factor is $\lvert \mathcal{B} \rvert ^2 = 4N^2(1-1/n^2)^2$ when dark photons occur at the center frequency $f_c$, where $n$ is the refractive index of dielectric disks.
	
	The material of the disks was chosen to be $\rm{LaAlO_3}$, with a refractive index $n \approx 5$. To construct a half-wave stack with the center frequency around $94\,\rm{GHz}$, the thickness $d_e$ and spacing $d_v$ of the disks were carefully designed as $0.320\, \rm{mm}$ and $1.580\, \rm{mm}$, respectively. The low dielectric loss angle $\delta_e$ of $\rm{LaAlO_3}$ \cite{sebastian_low-loss_2015} enhances the dark photon conversion rate by minimizing the energy loss of induced photons as they propagate through the disks. The surfaces of dielectric disks and mirror were carefully polished. The mirror was coated with a $2\,\rm{\upmu m}$ thick layer of gold, which is much larger than the skin depth of gold ($\sim 250\,\rm{nm}$). Notably, the half-wave stack exhibited robustness against imperfections in disk thickness and spacing arising from machining and installation errors.
	
	\begin{figure}[h]
		\includegraphics[width=1\linewidth]{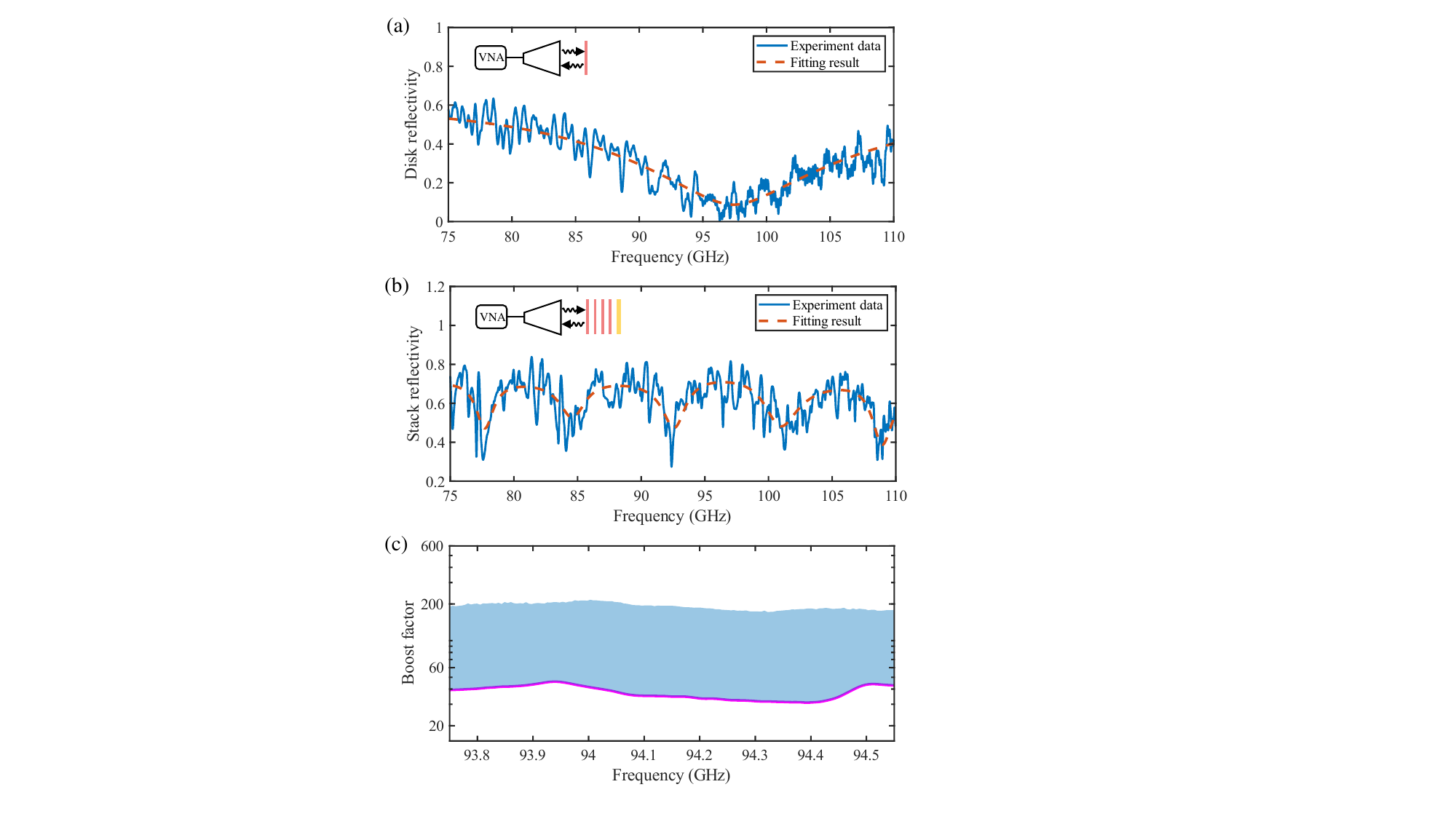}
		\caption{\label{fig:2} Results of boost factor calibration with the reflectivity test method. (a) The reflectivity of the first dielectric disk (the bottom one in the stack). The blue solid line is the measurement data, and the red dashed line is the fitting result with $\mathcal{R}$. The inset shows the scheme of the reflectivity test using a vector network analyzer (VNA). (b) The reflectivity of stack. The blue solid line and red dashed line stand for experiment data and the fitting result with the one-dimension multiple layers model, respectively. (c) Boost factor as a function of frequency. The light blue shaded region indicates the range of boost factor obtained by varying the stack and antenna parameters within their uncertainties. The magenta line shows the minimum value of boost factor.}
	\end{figure}
	
	A linearly polarized antenna was positioned above the dielectric stack to collect the converted photons, and it was sensitive to dark photon signals with the same polarization direction. The antenna and stack were placed in a metal shielding to isolate the environmental electromagnetic noise. After being coupled into the waveguide by the antenna, the dark photon signal entered the receiver chain for detection. The millimeter-wave low noise amplifier chain within the receiver chain amplified the dark photon signal by a total gain of $\sim 60\,\rm{dB}$. The amplified signal was subsequently down-converted to the baseband frequency below $205\,\rm{MHz}$ through multi-stage filtering and mixing. The millimeter-wave local oscillator provided a tunable local oscillating signal ranging from $93.7\,\rm{GHz}$ to $94.4\,\rm{GHz}$. To analyse the frequency domain properties of the signal, we employed a DAQ board (JYTEK, JY-9824) acting as a real-time spectrum analyzer with the bandwidth of $250\,\rm{MHz}$. To improve the dark photon search efficiency, the DAQ board was updated to support parallel data acquisition and Fourier transformation with negligible dead time. The dielectric stack, antenna and receiver chain were operated at room temperature. The added noise of the receiver chain varied from $(528\pm58)\,\rm K$ to $(937\pm102)\,\rm K$ in the frequency range of $93.750 – 94.550 \,\rm{GHz}$. There was a small input reflection of the receiver chain. This mismatch would modify the dark photon signal by a factor $F_{\rm{RC}}$, which was above $0.897$ in the searching frequency range (see more details in Appendix \cite{wei_supplementary_nodate}).
	
	The collected dark photon signal is proportional to the boost factor $\lvert \mathcal{B} \rvert^2$, which should be carefully characterized. The emitted signal power is determined by the stack configuration, including the thicknesses $d_e$, spacing $d_v$, refractive indices $n$ and loss angles $\delta$ of $\rm{LaAlO_3}$ disks \cite{millar_dielectric_2017}. The loss angle $\delta$ comprises two components: the intrinsic dielectric loss $\delta_e$ of the $\rm{LaAlO_3}$ material, and an additional loss arising from three-dimension effects—such as disk tilts within the stack—modeled as an effective loss angle $\delta_{\rm{eff}}$. The stack parameters above were accurately calibrated by carefully designed procedures. The thicknesses $d_e$ of the four $\rm{LaAlO_3}$ disks were $0.312\pm 0.007\,\rm{mm}$, $0.311\pm 0.007\,\rm{mm}$, $0.308\pm 0.006\,\rm{mm}$ and $0.319\pm 0.006\,\rm{mm}$ (from bottom to top in Fig.\,\ref{fig:1}(b)), which were obtained by a spiral micrometer. The rest stack parameters were calibrated by reflectivity tests with a millimeter-wave vector network analyzer (Ceyear, 3672C). Firstly, the reflectivity test of each $\rm{LaAlO_3}$ disk was implemented. The reflectivity of a single dielectric disk can be represented as \cite{millar_dielectric_2017}
	\begin{equation}
		\mathcal{R} = \eta \left| \frac{(\tilde{n}^2-1)\sin\Delta}{2i\tilde{n}\cos\Delta + (\tilde{n}^2+1)\sin\Delta}\right|,
		\label{eq:four}
	\end{equation}
	where $\eta <1$ results from the energy loss while antenna emitting and receiving the photons. $\Delta = \tilde{n}\omega d_e/c$ denotes the phase shift of photons when going through the disk, where $\omega$ is the frequency, and $\tilde{n}=ne^{i\delta_{\rm{single}}/2}$ is the complex refractive index including the loss angle $\delta_{\rm{single}}$ within the single disk test. The reflectivity of the first $\rm{LaAlO_3}$ disk (the bottom one in Fig.\,\ref{fig:1}(b)) is shown in Fig.\,\ref{fig:2}(a). By fitting the experiment result with $\mathcal{R}$, the refractive index $n$ was extracted as $4.93\pm 0.11$. Reflectivity tests of the other $\rm{LaAlO_3}$ disks resulted in $n$ as $4.89\pm 0.11$, $4.94\pm 0.10$ and $4.91\pm 0.09$, respectively (see Appendix \cite{wei_supplementary_nodate} for the reflectivity measurements of other disks). Then, the stack reflectivity test was implemented to obtain loss angles $\delta$ and spacing $d_v$. The result is shown in Fig.\,\ref{fig:2}(b). By fitting the experiment measurements with the one-dimension multiple layers model \cite{millar_dielectric_2017}, the spacing $d_v$ were given as $1.545\pm0.021\,\rm{mm}$, $1.569\pm0.015\,\rm{mm}$, $1.570\pm0.022\,\rm{mm}$, and $1.560\pm0.008\,\rm{mm}$, respectively. The loss angles $\delta$ were also obtained as $(9.89\pm2.57)\times10^{-3}$, $(5.30\pm2.94)\times10^{-3}$, $(1.14\pm0.32)\times10^{-2}$ and $(1.17\pm0.29)\times10^{-2}$, respectively. Due to the extra loss caused by three-dimension effects, the loss angles $\delta$ were larger than the $\rm{LaAlO_3}$ dielectric loss angle $\delta_e$, which is $\sim 3\times10^{-3}$ at room temperature\cite{alford_dielectric_2001}. 
	
	To characterize the boost factor $\lvert \mathcal{B} \rvert^2$, the imperfections of antenna also require careful analysis and calibration, which comprise the energy loss and the reflection of antenna. In order to analyze the antenna's imperfections, we constructed a one-dimension model of antenna, where the energy loss and the reflection were modeled by frequency-dependent parameters (see Appendix \cite{wei_supplementary_nodate} for details). Since these antenna parameters varied with frequency, it was difficult to fully extract them through the stack reflectivity fit in Fig.\,\ref{fig:2}(b). Instead, we implemented the antenna-mirror reflectivity tests to calibrate these parameters, where the distance between the mirror and antenna was scanned for about 2 mm. The antenna parameters were obtained through fitting the antenna model to these reflectivity measurements at different frequency points. Then, to characterize boost factor $\lvert \mathcal{B} \rvert^2$, we included the antenna model into the one-dimension multiple layers model. The minimum value of boost factor was acquired by scanning the parameters of stack and antenna within their uncertainties, as indicated by the magenta line in Fig.\,\ref{fig:2}(c). In the frequency range of $93.750 – 94.550 \,\rm{GHz}$, the minimum value of boost factor varied from $31$ to $46$. In the following analysis, we took this minimum value of $\lvert \mathcal{B} \rvert^2$ as conservative estimates to set constraints on the kinetic mixing $\chi$. 
	
	The data acquisition was conducted during the period from 2025-01-13 to 2025-01-22, across eight overlapping frequency ranges, collectively covering a range from $93.750$ to $94.550 \,\rm{GHz}$ (see more details of the data acquisition in Appendix \cite{wei_supplementary_nodate}). The data acquisition time of each frequency range was about 24 hours. Within the total acquisition time of $t_{\rm{tot}}\approx7\times10^5\, \rm s$, a total number of $N_{\rm{tot}}\approx5\times10^9$ spectra were acquired. The frequency bin width $\delta f$ of each spectrum was $7.63\,\rm{kHz}$. The spectra were  averaged on the DAQ hardware of $5.1\times10^3$ times before being saved, resulting in around $1\times10^6$ recorded spectra.
	
	\begin{figure}[h]
		\includegraphics[width=1\linewidth]{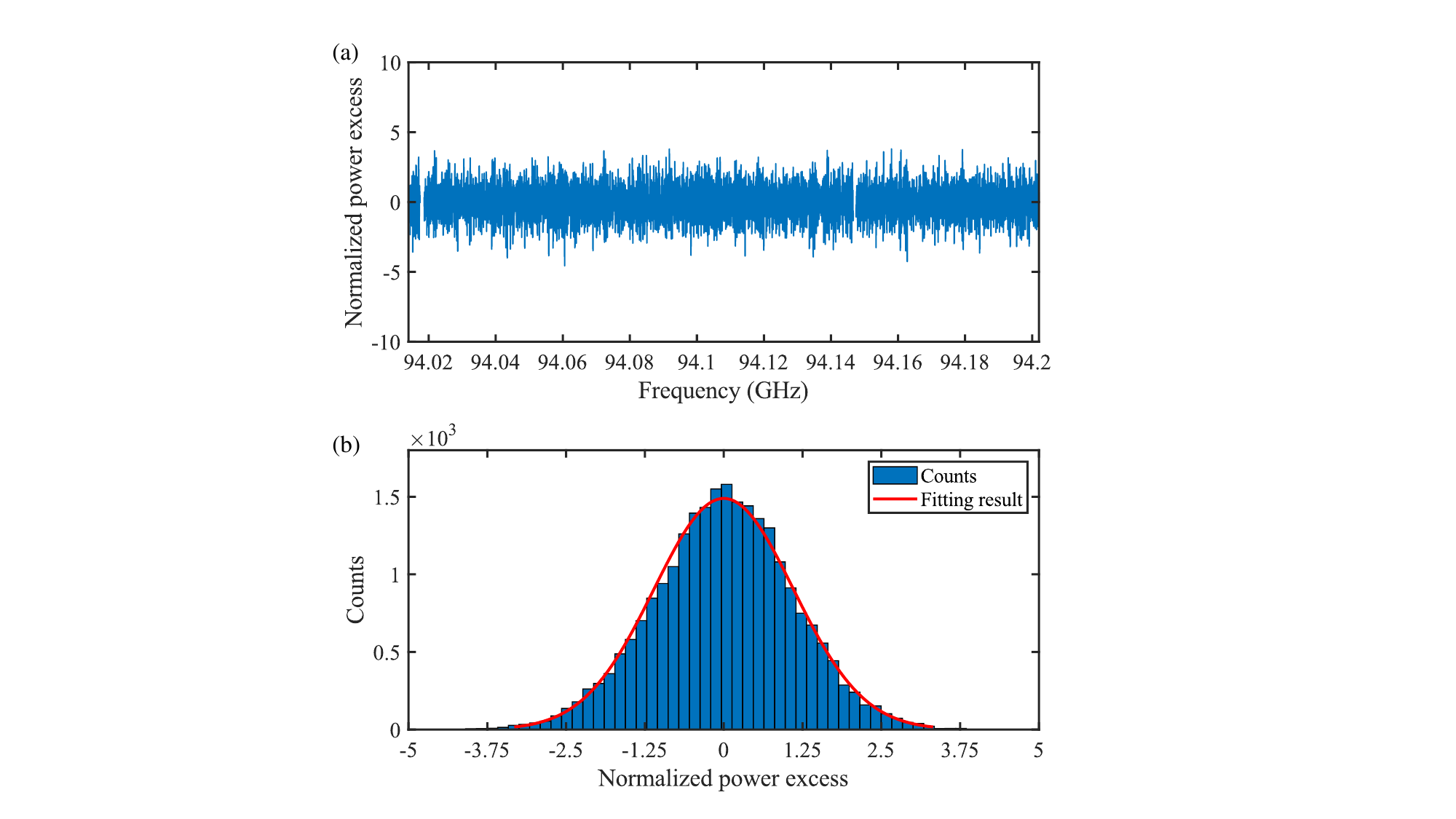}
		\caption{\label{fig:3} Data analysis of the frequency range from $94.014 \,\rm{GHz}$ to $94.202 \,\rm{GHz}$. (a) The normalized power excess after convolution. (b) The normalized power excess distribution. The blue bars show the histogram of normalized power excess, and the red solid line is the fitting result with normal Gaussian distribution.}
	\end{figure}
	
	The data analysis procedures of the eight frequency ranges were the same and here we take the frequency range from $94.014 \,\rm{GHz}$ to $94.202 \,\rm{GHz}$ as an example. The number of the saved spectra was $124010$, and they were averaged to a single raw spectrum. The raw power spectrum was filtered by a 4th-order Savitzky-Golay (SG) filter with the window length of $1.14 \,\rm{MHz}$ \cite{cervantes_admx-orpheus_2022}. Through dividing the raw power spectrum by the baseline obtained from the SG filter, we transformed it into the raw power excess. This baseline removal operation would attenuate a potential dark photon signal by a factor of $\eta_{\rm SG} = 0.74$, which was obtained from analysis of a simulated dark photon signal. Since the expected dark photon signal would occupy tens of bins, the raw power excess was convolved with the dark photon line shape to enhance the SNR \cite{cervantes_admx-orpheus_2022}. The standard deviation of convolved power excess was also obtained from convolution. Then, the convolved power excess was divided by its standard deviation to get the normalized power excess, as shown in Fig.\,\ref{fig:3}(a). There were no bins exceeding $5\sigma$ in the normalized power excess, and it satisfied the normal Gaussian distribution $\rm{N(\mu,\sigma^2)}$. The parameters of mean $\mu=0.000\pm0.014$ and standard deviation \rm{$\sigma=1.108\pm0.010$} were obtained from the fitting result, as illustrated by the red line in Fig.\,\ref{fig:3}(b). The standard deviation $\sigma$ was slightly larger than one due to the correlations induced by the SG filter and the Hanning window while performing FFT.
	
	\begin{figure*}
		\includegraphics[width=1\linewidth]{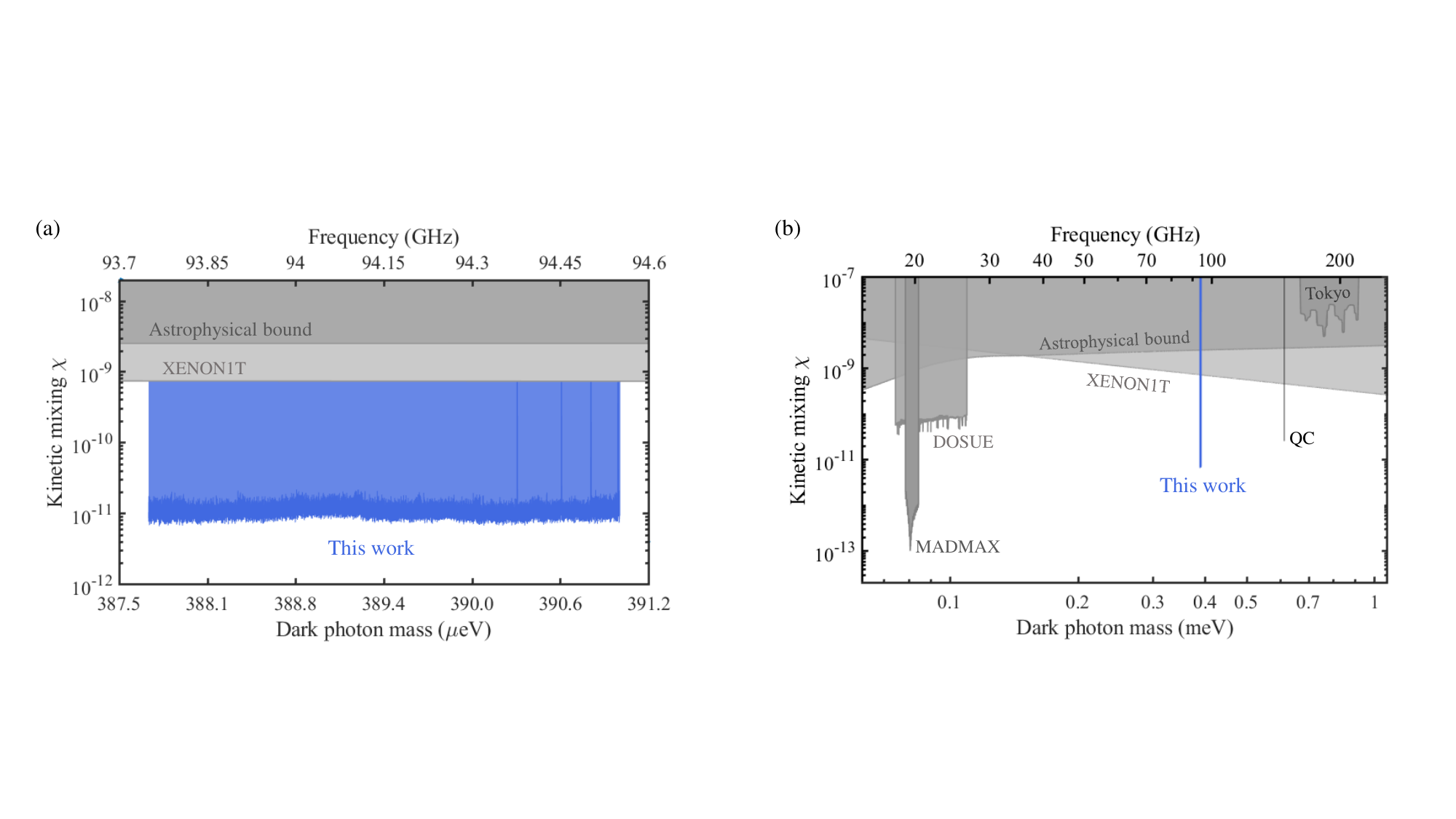}
		\caption{\label{fig:4}  Constraints on the dark photon with kinetic mixing $\chi$ and dark photon mass. (a) The blue shaded region shows the $90\%$ confidence limits set by this work. The light gray region refers to the result of XENON1T \cite{aprile_emission_2022}, and the deep gray region is the astrophysical bound set by Arias et al \cite{arias_wispy_2012}. (b) This work along with other experimental searches in an extended dark photon mass range. Constraints on kinetic mixing from MADMAX \cite{madmax_collaboration_first_2025}, DOSUE \cite{kotaka_search_2023}, QC \cite{fan_one-electron_2022} and Tokyo \cite{knirck_first_2018} are shown in gray.}
	\end{figure*}
	
	In the raw power excess spectra of eight frequency ranges, there were 20 candidate peaks sticking out. These peaks were ruled out by the overlapping data in neighboring frequency ranges or by the blank tests (see Appendix \cite{wei_supplementary_nodate} for details on the rejection of candidate peaks). Our results gave non observation of any possible dark photon signal within our detection frequency range. Based on that, the $90\%$ confidence constraints on kinetic mixing $\chi$ were given by Bayesian analysis, which is developed in the ADMX experiment \cite{cervantes_admx-orpheus_2022}. The results of eight frequency ranges were combined by selecting the more stringent constraints at the overlapping frequency bins, as shown in Fig.\,\ref{fig:4}(a). We established the most stringent constraints that $\chi < 2.2\times 10^{-11}$ in the dark photon mass region from $387.72\,\rm{\upmu eV}$ to $391.03\,\rm{\upmu eV}$. An upper limit of $\chi < 6.68\times 10^{-12}$ was achieved at $387.79\,\rm{\upmu eV}$, which improves the existing limits \cite{aprile_emission_2022} by about 110 times. Constraints from other haloscope searches in an extended mass range are shown in Fig.\,\ref{fig:4}(b). This work establishes the most stringent constraints within the W band, and notably, the parameter space region explored by this work is well-motivated under different dark photon production mechanisms (see Appendix \cite{wei_supplementary_nodate} for the discussion of dark photon production mechanisms). 
	
	In summary, we constructed the first millimeter-wave dielectric haloscope and calibrated the boost factor. Based on this dielectric haloscope, we launched a wide-band search for randomly polarized dark photon dark matter. In the range of dark photon masses from $387.72\,\rm{\upmu eV}$ to $391.03\,\rm{\upmu eV}$, we achieved the most stringent constraints on kinetic mixing, which improved the existing limits by two orders of magnitude. This work provided essential techniques for the implementation of dielectric haloscope in the millimeter-wave range. In the future, by introducing cryogenic amplifiers with lower noise, such as SIS mixer, the constraints could be further improved by a factor of $\sim 3$. By enhancing the millimeter-wave local oscillator, we could explore a wider dark photon mass range from $380.48\,\rm{\upmu eV}$ to $417.70\,\rm{\upmu eV}$ with the same stack. Furthermore, our dielectric haloscope could also be utilized to search for axion dark matter, if the stack is placed in a steady magnetic field which is parallel to the dielectric disks. By employing a narrow-band yet more sensitive stack configuration and increasing the number of disks, this approach has the potential to set new constraints on the axion-photon coupling parameter within the millimeter-wave frequency range.
	
	\begin{acknowledgments}
		This work was supported by NSFC (T2388102, 12205290, 12261160569), the Innovation Program for Quantum Science and Technology (2021ZD0302200), and the National Key R$\&$D Program of China (Grant No. 2021YFC2203100). X.R. thanks the support by the Major Frontier Research Project of the University of Science and Technology of China (Grant No. LS9990000002). M.J. thanks the Fundamental Research Funds for Central Universities.
	\end{acknowledgments}
	
	
	%
	
	\appendix
	\section{\label{sec:1} The schematic diagram of the experimental apparatus}
\begin{figure*}
		\includegraphics[width=1\linewidth]{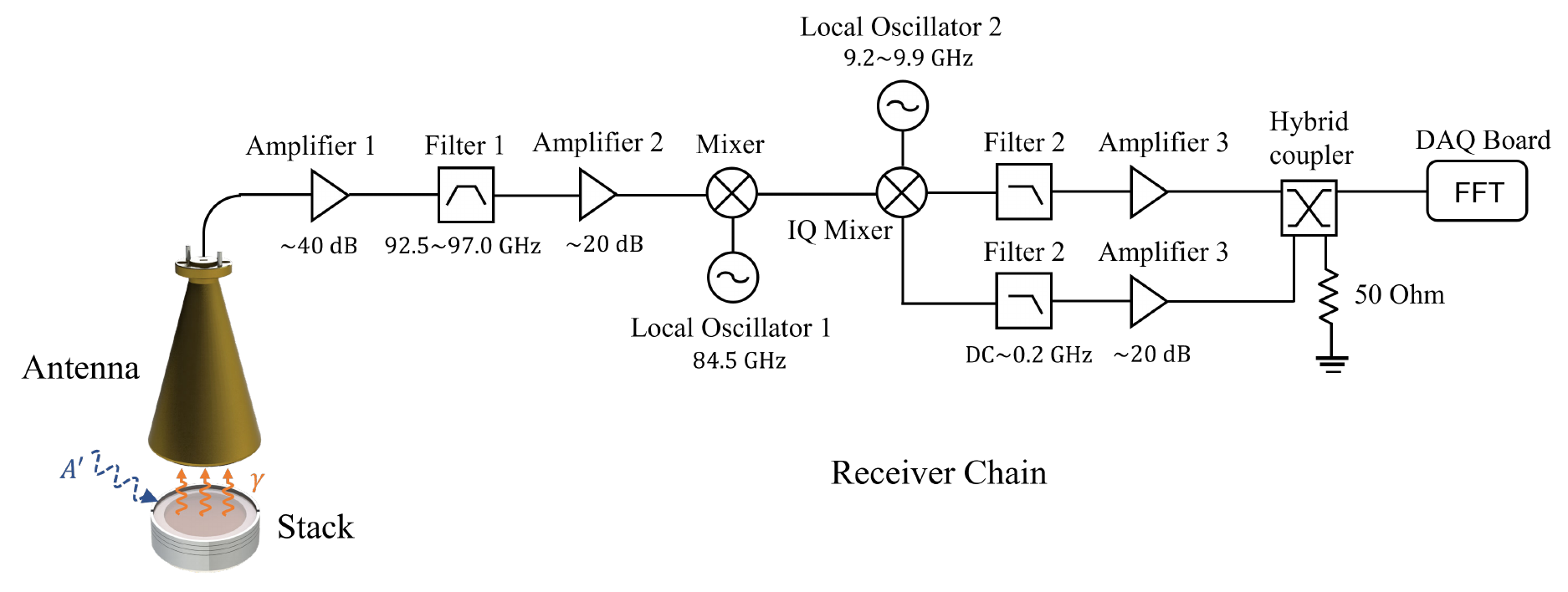}
		\caption{\label{fig:S1} Schematic diagram of the experiment apparatus for dark photon search.}
	\end{figure*}

	The schematic diagram of the experimental apparatus is shown in Fig.\,\ref{fig:S1}, where more details of the receiver chain is given. Amplifier 1 and  Amplifier 2 were millimeter-wave low noise amplifiers working in the range of $65–110\,\rm{GHz}$, which provided a total gain of around $60\,\rm{dB}$. The input noise of Amplifier 1 was the predominant source of the receiver chain noise. The down-conversion was achieved by a two-step frequency mixing. Local oscillator 1 supplied a fixed local oscillating signal of $84.5\,\rm{GHz}$. The Filter 1 was introduced to suppress the image-frequency noise of Amplifier 1 during the first-stage frequency mixing. Local oscillator 2 offered a tuning local oscillating signal ranging from $9.2\,\rm{GHz}$ to $9.9\,\rm{GHz}$, enabling a dark photon search bandwidth of $0.8\,\rm{GHz}$. An IQ mixer, in conjunction with a 90-degree hybrid coupler, was employed to suppress image-frequency noise during the second-stage mixing. The down-converted signal at baseband frequency ($<205\,\rm{MHz}$) was finally acquired by an FPGA-based DAQ board, which was equipped with $1\,\rm{GHz}$ sampling rate and $250\,\rm{MHz}$ bandwidth.
	
	\section{\label{sec:2} The reflectivity tests}
\begin{figure}[h]
		\includegraphics[width=1\linewidth]{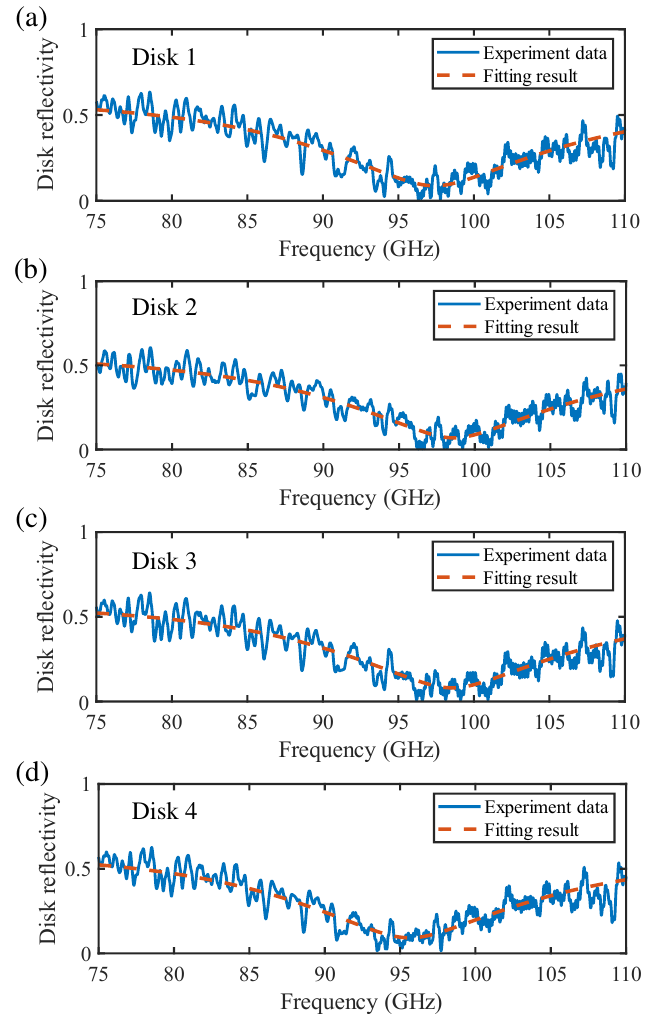}
		\caption{\label{fig:S2} The reflectivity tests of the rest dielectric disks. (a)–(d) The reflectivity and fitting results of Disk 1–4.}
	\end{figure}

	The reflectivity test results of four $\rm{LaAlO_3}$ disks are shown in Fig.\,\ref{fig:S2}. Disk 1–4 denote the four disks from bottom to top in the stack. The loss angles $\delta_{\rm{single}}$ obtained from the single disk tests were $(4.24\pm0.43)\times10^{-2}$, $(3.50\pm0.42)\times10^{-2}$, $(4.01\pm0.44)\times10^{-2}$ and $(4.49\pm0.42)\times10^{-2}$, respectively. The results of $\delta_{\rm{single}}$ are different from the stack loss angles $\delta$. This is due to the installation differences between the single-disk and stack tests, which resulted in distinct three-dimensional effects. 
	
\begin{figure}[h]
		\includegraphics[width=1\linewidth]{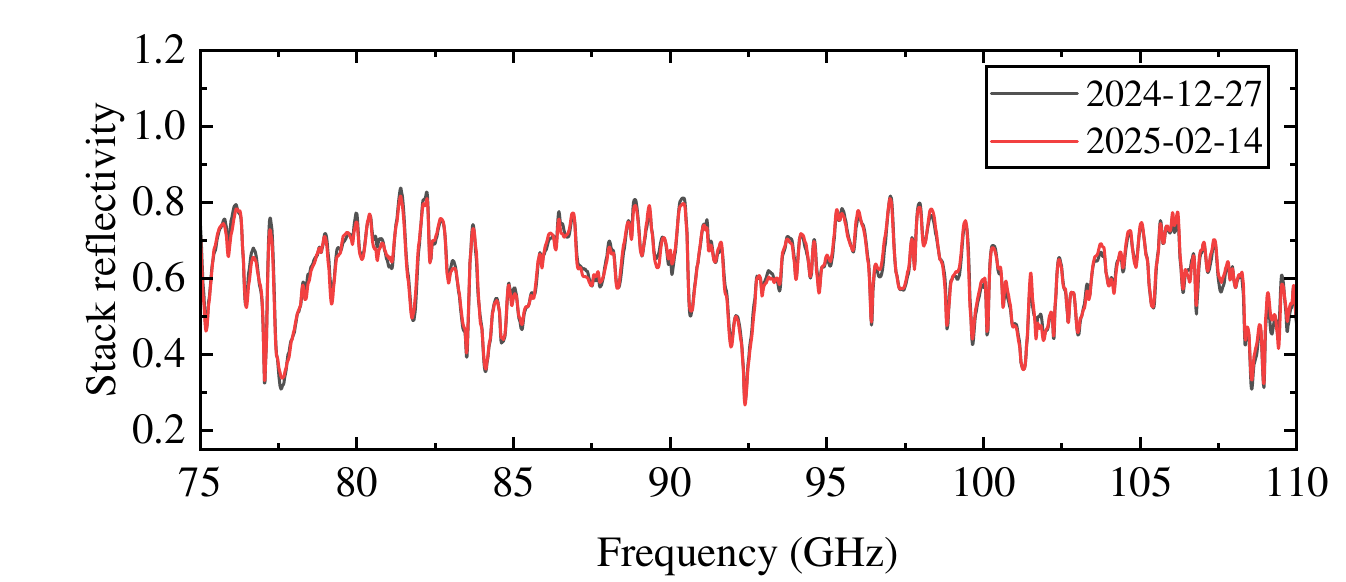}
		\caption{\label{fig:S3} The reflectivity tests of the stack at 2024-12-27 and 2025-2-14, respectively.}
	\end{figure}

	We conducted stack reflectivity tests at 2024-12-27 and 2025-02-14, separately. The results demonstrated that the stack remained quite stable within a long period, as shown in Fig.\,\ref{fig:S3}. The fitting results of the two tests yielded a relative deviation in the boost factor of $\sim 0.1\%$, which was negligible compared with other system uncertainties.
	
	\section{\label{sec:3} Calibration of the antenna}
\begin{figure}
		\includegraphics[width=1\linewidth]{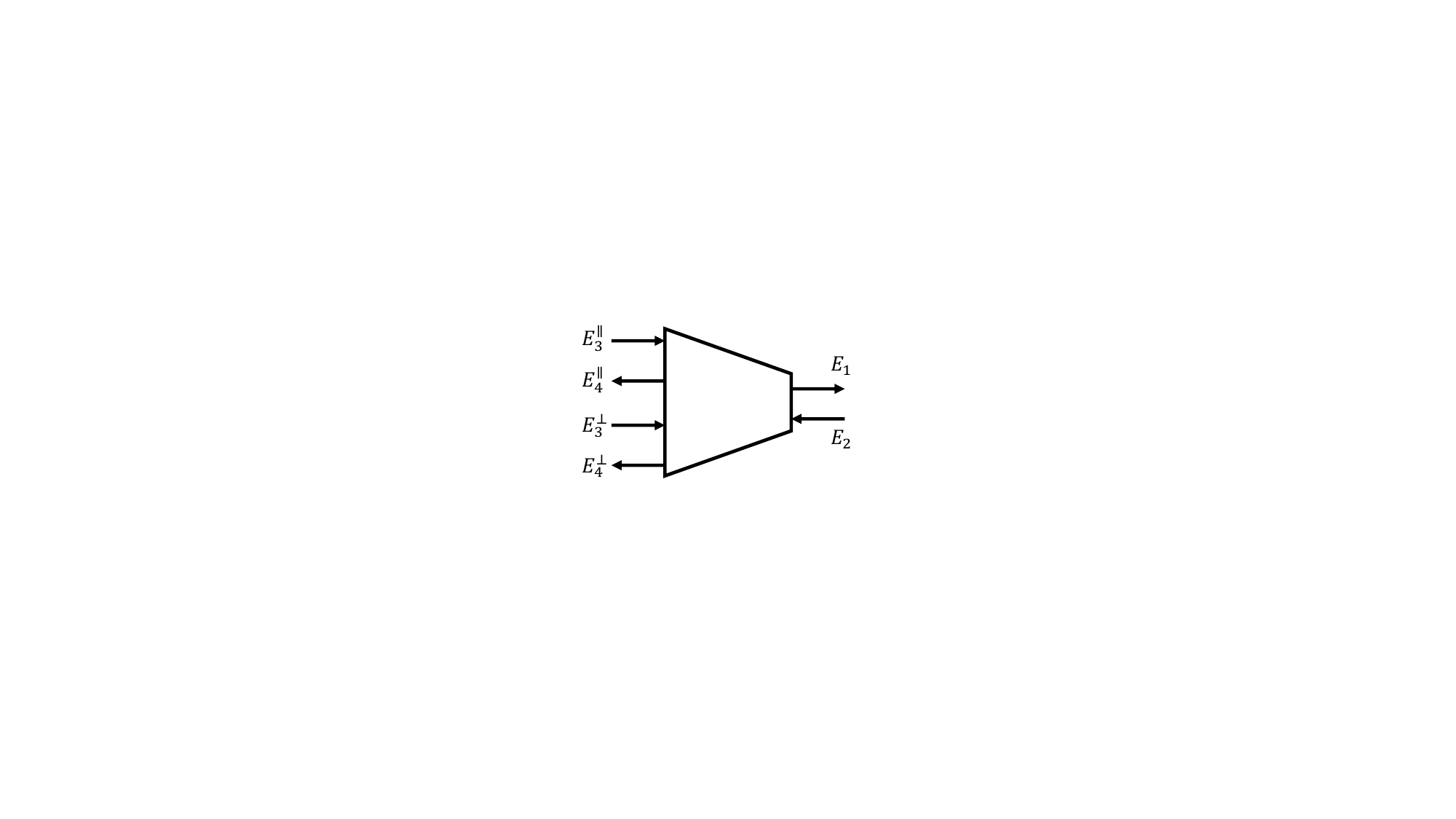}
		\caption{\label{fig:S4} The one-dimension model of antenna. The solid straight arrows indicate the electric fields of left and right moving electromagnetic waves. $E_3^{\parallel}$ and $E_4^{\parallel}$ are the electric fields with the same polarization as the antenna, and $E_3^{\perp}$ and $E_4^{\perp}$ are the electric fields with perpendicular polarization.}
	\end{figure}

\begin{figure}[h]
		\includegraphics[width=1\linewidth]{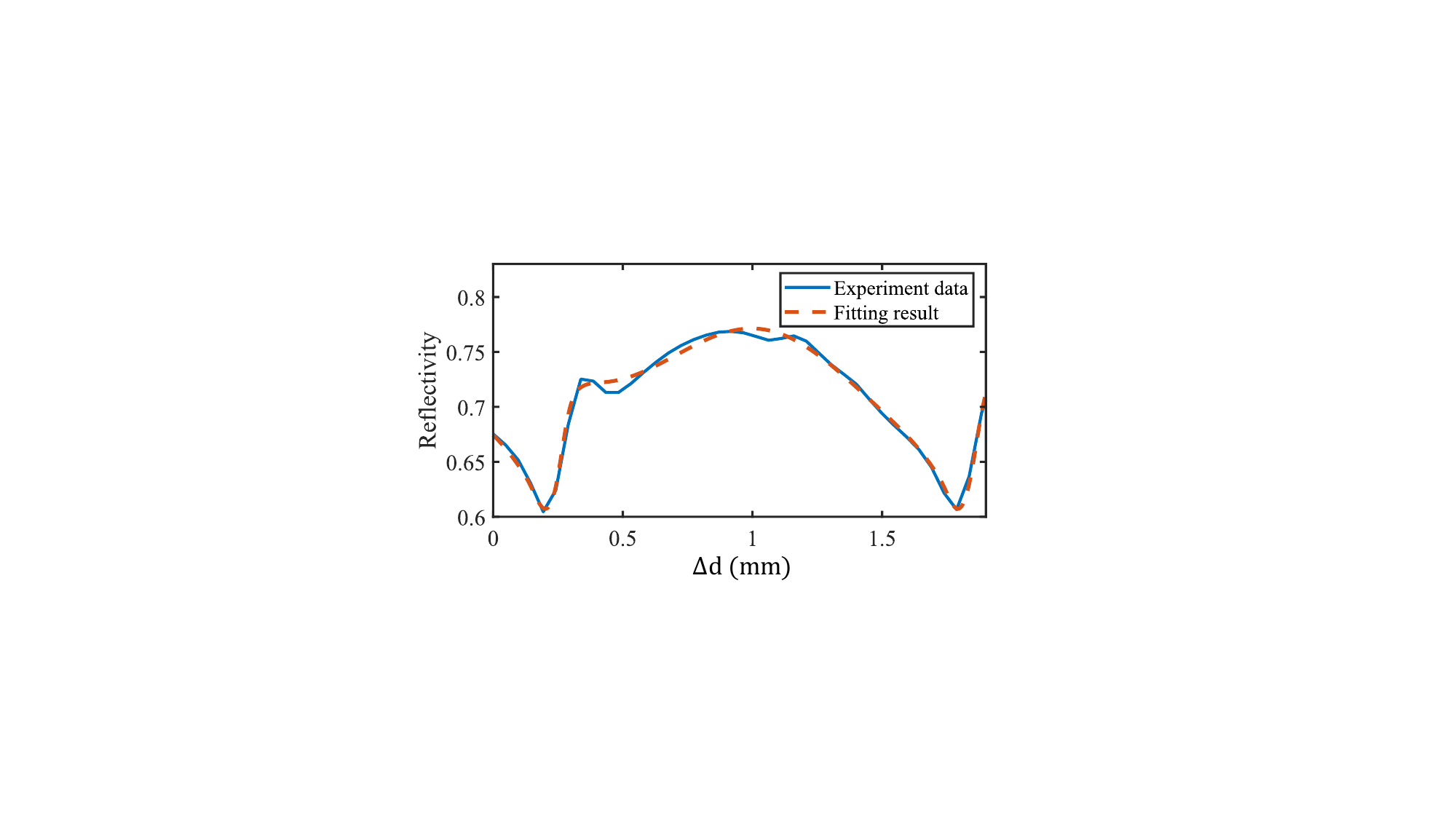}
		\caption{\label{fig:S4_sub2} The antenna-mirror S11 test at $94.130\,\rm{GHz}$, where $\Delta \rm d$ is the displacement of the mirror. The blue solid line represents the measurements, and the red dashed line shows the fitting result with the one-dimension antenna model.}
	\end{figure}

\begin{figure}[h]
		\includegraphics[width=1\linewidth]{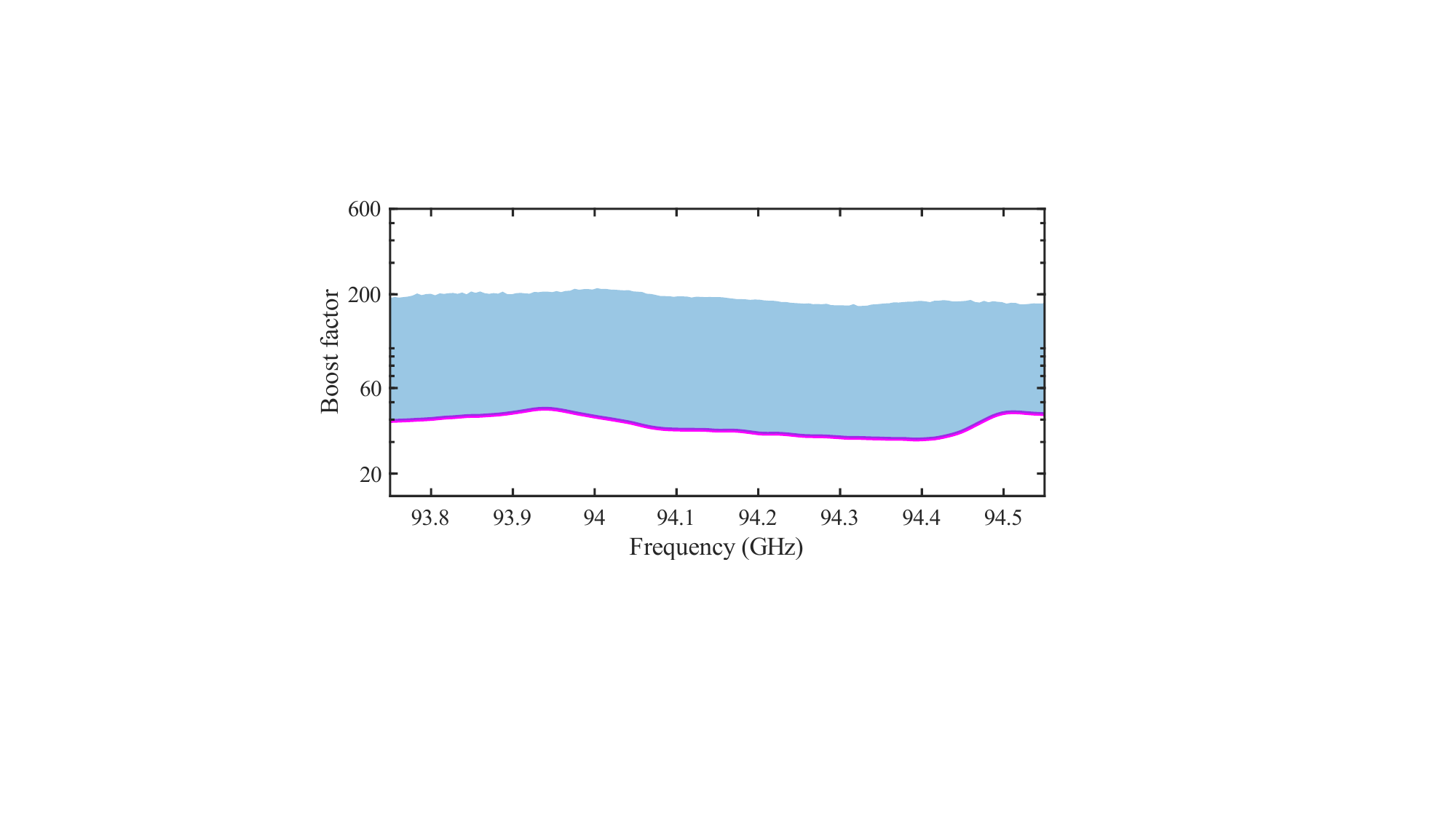}
		\caption{\label{fig:S4_sub3} The results of boost factor. The light blue shaded region indicates the range of boost factor obtained by varying the parameters within uncertainties. The magenta line shows the minimum value of boost factor.}
	\end{figure}

	There exists the energy loss when the antenna receiving the dark photon signal. The reflection of the antenna will induce the standing wave between the stack and antenna, which will also modify the dark photon signal. In order to analyze these imperfections of antenna, we constructed a one-dimension model of antenna, as shown in Fig.\,\ref{fig:S4}. The signal with polarization parallel to the antenna is denoted as $E_3^{\parallel}$, and the signal with perpendicular polarization is represented by $E_3^{\perp}$. The reflections of antenna are denoted as $E_4^{\parallel}$ and $E_4^{\perp}$, respectively. Notably, upon reflection from the antenna, a small portion of the signal undergoes polarization conversion, resulting in a component with orthogonal polarization. Therefore, this model can be described as follows,
	\begin{align}
		E_1 &= \eta_A E_3^{\parallel} + \eta_A' E_3^{\perp} + \Gamma_r E_2, \\
		E_4^{\parallel} &= \Gamma_A E_3^{\parallel} + \Gamma_t E_3^{\perp} + \eta_A E_2, \\
		E_4^{\perp} &= \Gamma_A' E_3^{\perp} + \Gamma_t E_3^{\parallel} + \eta_A' E_2.
	\end{align}
	For the signal with parallel polarization, the antenna transmission efficiency is denoted as $\eta_A$, and for the perpendicular polarization it is $\eta_A'$. The antenna reflectivity for the parallel-polarization and perpendicular-polarization signals is denoted as $\Gamma_A$ and $\Gamma_A'$, respectively. The $\Gamma_r$ represents the return loss of antenna, and $\Gamma_t$ represents the transformation between signals of different polarization directions upon reflection from the antenna. Notably, these  parameters of antenna are frequency-dependent. Since the antenna is linearly polarized, the transmission efficiency $\eta_A'$ for the perpendicular-polarized signal is set as zero. The typical energy return loss of the antenna is low as $-25\,\rm{dB}$, so $\Gamma_r$ can also be neglected. Therefore, the model can be reduced as
	\begin{align}
		E_1 &= \eta_A E_3^{\parallel}, \\
		E_4^{\parallel} &= \Gamma_A E_3^{\parallel} + \Gamma_t E_3^{\perp} + \eta_A E_2, \\
		E_4^{\perp} &= \Gamma_A' E_3^{\perp} + \Gamma_t E_3^{\parallel}.
	\end{align}
	In order to calibrate the frequency-dependent antenna parameters of $\eta_A$, $\Gamma_A$, $\Gamma_A'$ and $\Gamma_t$, the antenna-mirror reflectivity tests were implemented, where we measured the reflectivity only with the mirror and without the disks installed. The distance between the mirror and the antenna was varied by around $2\,\rm{mm}$, and the results were fitted with this one-dimension model to obtain the antenna parameters as well as their uncertainties. An example at $94.130\,\rm{GHz}$ is shown in Fig.\,\ref{fig:S4_sub2}.
	Then, the antenna model was included into the one-dimension multiple layers model of stack, and the boost factor was obtained by scanning the parameters of stack and antenna within their uncertainties. The results of boost factor is shown in Fig.\,\ref{fig:S4_sub3}. Notably, in the calculation of boost factor $\lvert \mathcal{B} \rvert ^2$ in Fig.\,\ref{fig:S4_sub3}, only the parallel-polarization component of dark photons was considered, so this $\lvert \mathcal{B} \rvert ^2$ can also be noted as $\lvert \mathcal{B_{\parallel}} \rvert ^2$. In fact, because of the polarization conversion during the antenna reflection, dark photons with perpendicular polarization can also generate a parallel-polarization signal, with the corresponding boost factor denoted as $\lvert\mathcal{B_{\perp}}\rvert^2$. However, within the frequency range from $93.750\,\rm{GHz}$ to $94.550\,\rm{GHz}$, the ratio $\lvert \mathcal{B_{\perp}/B_{\parallel}} \rvert ^2$ was found to be below $4\%$, indicating that the contribution from the perpendicular component of dark photons can be neglected.
	
	\section{\label{sec:4} Characterization of the receiver chain}
	In this work, we conducted the dark photon search across eight overlapping frequency ranges (ranges 1–8; see section\,\ref{sec:5}), by changing the local oscillator 2 from $9.2\,\rm{GHz}$ to $9.9\,\rm{GHz}$ with a step of $100\,\rm{MHz}$. This corresponded to a total local oscillator frequency spanning from $93.7\,\rm{GHz}$ to $94.4\,\rm{GHz}$. The characterization of the receiver chain was also implemented at those eight frequency ranges, separately.
	
\begin{figure}[h]
		\includegraphics[width=1\linewidth]{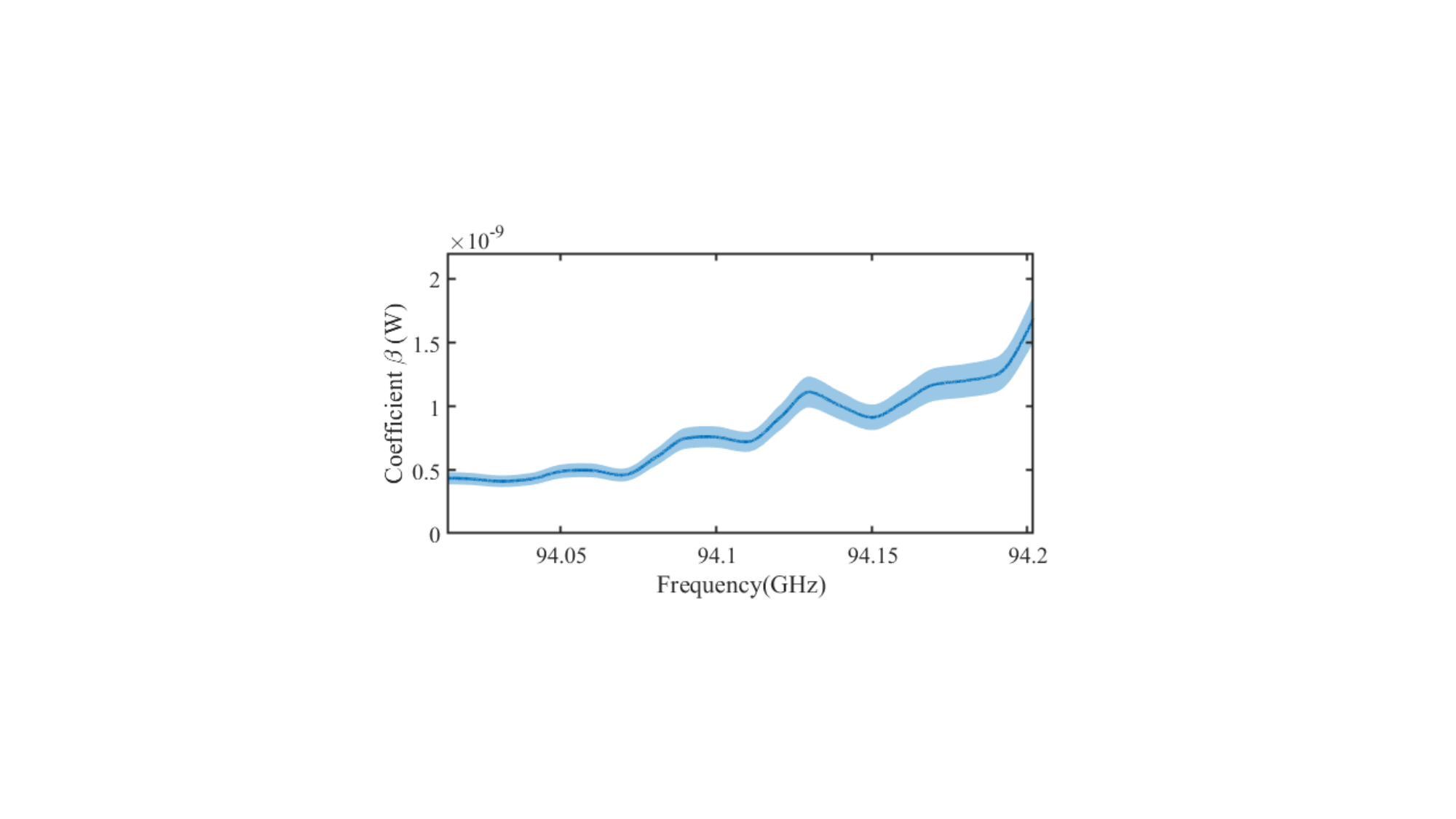}
		\caption{\label{fig:S5} The calibrated coefficient $\beta$ of frequency range 4, which is from $94.014 \,\rm{GHz}$ to $94.202 \,\rm{GHz}$. The shaded region denotes the $\pm1\sigma$ uncertainty of $\beta$.}
	\end{figure}

	To map the output of DAQ board to the input power of the receiver chain, the coefficient $\beta$ was calibrated first. The definition of this coefficient was $\beta = P_r/y$, where $P_r$ was the signal power that entered the receiver chain and $y$ was the dimensionless signal value recorded by DAQ, proportional to the signal power. By inputting a calibration signal with known power at the input of Amplifier 1, we measured the DAQ output $y$ at different frequencies. The coefficient result in the frequency range of $94.014–94.202 \,\rm{GHz}$ (frequency range 4) is shown in Fig.\,\ref{fig:S5}, the shaded region denotes the uncertainty ($\pm1\sigma$) of $\beta$. The uncertainty of $\beta$ consisted of the uncertainties of calibration signal power and DAQ output value.
	During the DAQ output measurement, the local oscillator frequency $f_{\rm{LO}}$ was calibrated at the same time. The calibration signal and DAQ board was triggered by one atomic clock, and $f_{\rm{LO}}$ was given by the difference between the signal frequency and the baseband frequency. The frequency stability test of the receiver chain was implemented by monitoring the baseband frequency of signal on the DAQ board. During a 24-hour monitoring period, the variation range of the baseband frequency was $22.8\,\rm{kHz}$. The fluctuation of the receiver chain frequency would cause an attenuation of the dark photon signal. By simulation on a hypothetical dark photon signal, the attenuation factor $\eta_{f}$ was $1.00\,(0.97–1.00)$, the first quartile 0.97 was selected as a conservative estimation of $\eta_{f}$.
	
	The receiver chain had an impedance mismatch to the antenna, due to the small input reflection of Amplifier 1 as $\lvert\mathcal{R_{\rm{RC}}}\rvert\approx 0.074$. This small reflection would modify the dark photon signal power by a factor $F_{\rm{RC}}=[(1-\lvert \mathcal{R_{\rm{RC}}} \rvert^2)/(\lvert 1-\mathcal{R_{\rm{s}}}\mathcal{R_{\rm{RC}}} \rvert^2)]$ , where $\mathcal{R_{\rm{s}}}$ is the reflectivity of stack \cite{madmax_collaboration_first_2025}. The lower bound of $F_{\rm{RC}}$ was adopted as the correction to the dark photon signal, which was $F_{\rm{RC}}^{\rm{min}}=[(1-\lvert \mathcal{R_{\rm{RC}}} \rvert^2)/(1+\lvert \mathcal{R_{\rm{s}}} \rvert\lvert \mathcal{R_{\rm{RC}}} \rvert)^2]$, varying from 0.897 to 0.912 in the frequency range of $93.750–94.550\,\rm{GHz}$. In the following analysis, $F_{\rm{RC}}^{\rm{min}}$ was absorbed in the calculation of constraints on the kinetic mixing.
	
	\section{\label{sec:5} Details of data analysis}
	
	\subsection{\label{sec:5_1}Candidate peaks analysis}
\begin{figure*}
		\includegraphics[width=1\linewidth]{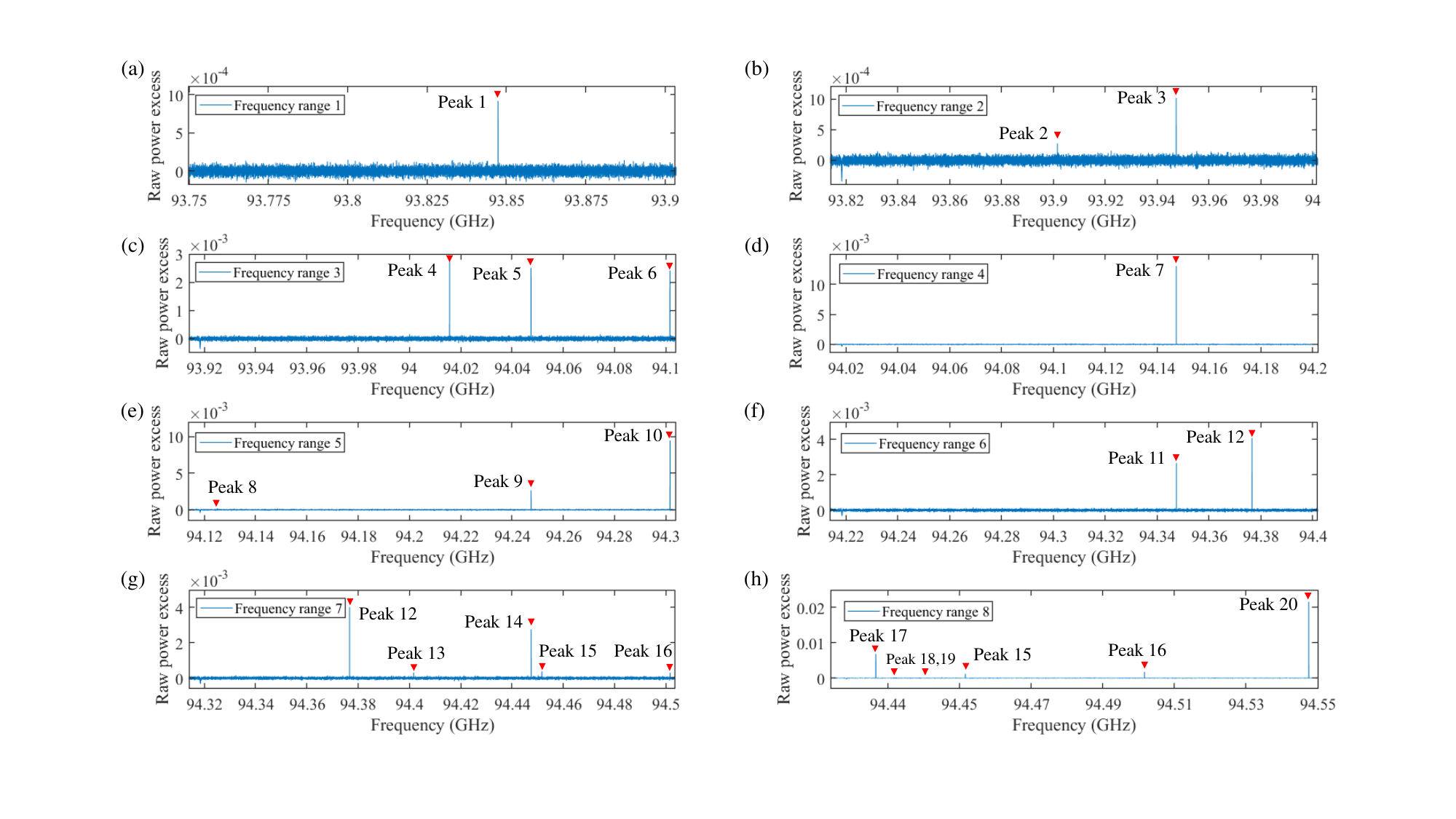}
		\caption{\label{fig:S6} The raw power excess results $\Delta_{\rm{raw}}$ of eight frequency ranges, which are obtained after the averaging and baseline-removal procedure of raw spectra. There are a total of twenty suspicious peaks. Peak 1–11, 13, 14, 17–19 do not exist in the result of the neighboring frequency range, and the remaining peaks are excluded through blank tests.}
	\end{figure*}

	We conducted the dark photon search across eight overlapping frequency ranges, and the data acquisition time was about 24 hours at each frequency range. After the averaging and baseline-removal procedure (see more details in Subsection \ref{sec:5_2}), twenty candidate peaks (Peak 1 to Peak 20) at different frequencies in the raw power excess spectra were discovered, as shown in Fig.\,\ref{fig:S6}. Sixteen of the candidate peaks (Peaks 1–11, 13, 14, 17–19) were not found in the overlapping region of neighboring frequency ranges, thus they were identified as artificial signals and ruled out. As for the rest four candidates (Peaks 12, 15, 16, 20), they were ruled out through the blank tests. We replaced the antenna and stack with a fixed load at Amplifier 1's input port. Then, we conducted 24-hour data acquisition in frequency range 7 and 8, and obtained about 600 million spectra respectively. The average and baseline removal procedures were conducted to obtain the raw power excess spectra. In the raw power excess spectra of blank tests, peaks 12,15,16 and 20 were also observed. The blank tests showed that the peaks 12,15,16 and 20 were artificial signals that came from the receiver chain, such as the amplifiers, local oscillators or DAQ board. The bins around these 20 peaks were excluded from analysis and thus from the constraints of every single frequency range. In the final constraints that combining the results of eight frequency ranges, we did not give constraints at the frequencies around peaks 12,15,16 and 20 due to these chronic artificial signals.
	
	Besides the peaks mentioned above, there were seven deeps in the results of frequency range 2–8. The deeps all had a similar linewidth around $200\,\rm{kHz}$, and all occurred at the same baseband frequency around $18\,\rm{MHz}$. Those deeps, also artificial signals, were induced by the DAQ board or Amplifier 3.
	
	\subsection{\label{sec:5_2}Details of data analysis procedures}
	In this subsection, more details of the data analysis procedures are provided. 
	\begin{figure}
		\includegraphics[width=1\linewidth]{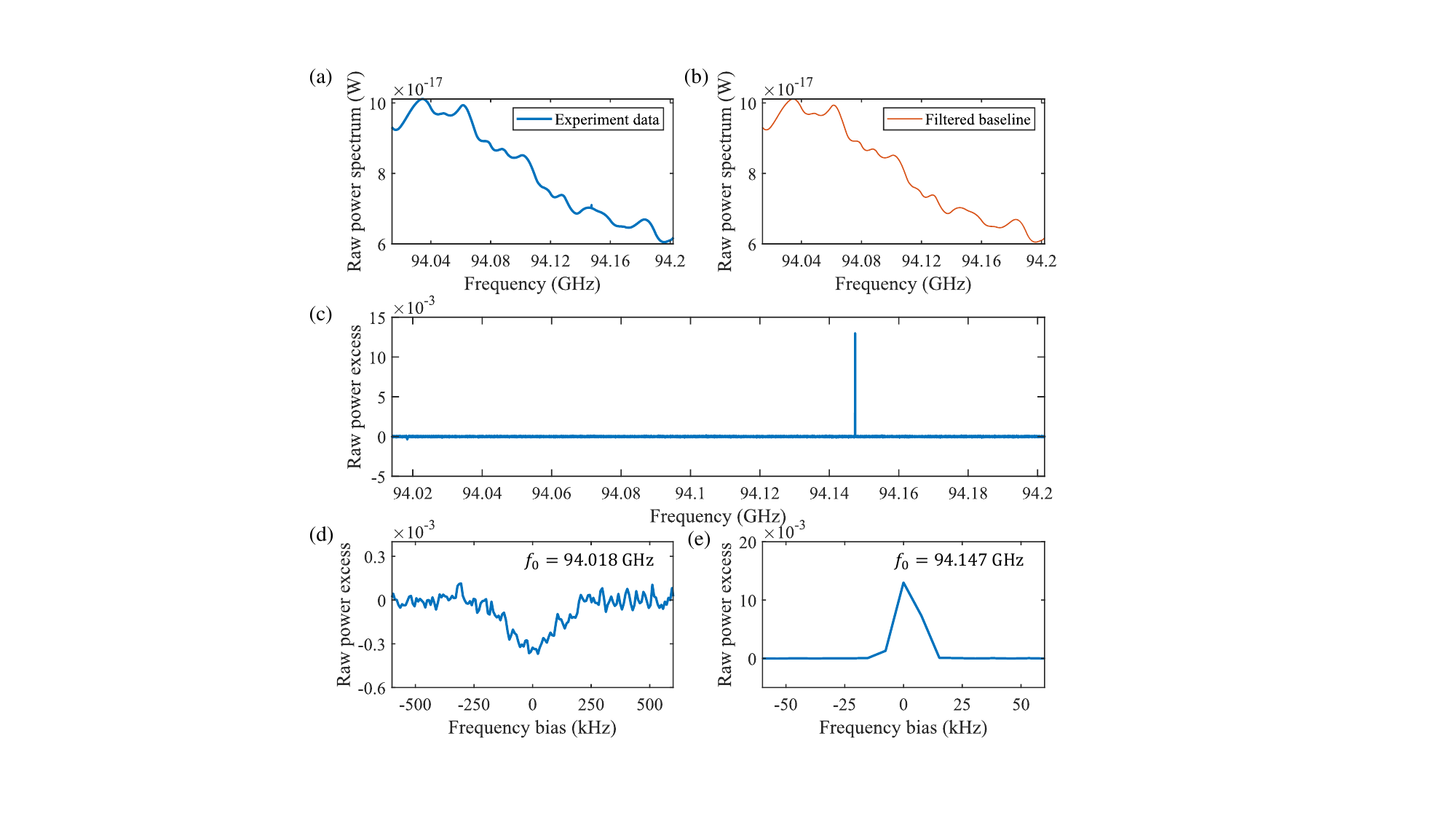}
		\caption{\label{fig:S7} (a) The experiment data of raw power spectrum $P_{\rm{raw}}$. (b) The baseline obtained from the SG filter. (c) The raw power excess $\Delta_{\rm{raw}}$ after the baseline removal operation and before excluding the artificial signals. (d)–(e) Zoomed-in view of two artificial signals in frequency range 4.}
	\end{figure}
	During the data acquisition time of eight days, a total of $\sim 5\times10^9$ spectra were acquired, and there were around $6\times10^8$ spectra of each frequency range. The acquired spectra were averaged by 5100 times on DAQ board and then saved. The frequency bin width $\delta f$ of each spectrum was set as $7.63\,\rm{kHz}$. Here we took the data of $94.014 –94.202 \,\rm{GHz}$ (frequency range 4) as an example. The saved 124010 spectra were averaged to a single raw spectrum $P'_{\rm{raw}}$. The raw spectrum $P'_{\rm{raw}}$ was multiplied by the coefficient $\beta$ to obtain the raw power spectrum $P_{\rm{raw}}$, which is shown in Fig.\,\ref{fig:S7}(a). Then, the raw power spectrum $P_{\rm{raw}}$ was filtered by a 4th-order Savitzky-Golay (SG) filter with a window length of $1.14\,\rm{MHz}$  \cite{cervantes_admx-orpheus_2022}. Notably, the SG filter would attenuate a potential dark photon signal by $\eta_{\rm{SG}}=0.74$. The value of $\eta_{\rm{SG}}$ was obtained by comparing the simulated dark photon signal amplitude before and after applying the SG filter. The filtered baseline is shown as the red line in Fig.\,\ref{fig:S7}(b), which was the average noise power $P_n$ of the system during the data acquisition time. The system noise fluctuated slightly ($\sim 1\%$) during the data acquisition time, which was considered as the system uncertainty of $P_n$. Then we obtained the raw power excess $\Delta_{\rm{raw}}$ as 
	\begin{equation}
		\Delta_{\rm{raw}} = \frac{P_{\rm{raw}}}{P_n}-1 ,
		\label{eq:S1}
	\end{equation}
	The raw power excess $\Delta_{\rm{raw}}$ is shown in Fig.\,\ref{fig:S7}(c), and the two artificial signals in this frequency range are shown in Fig.\,\ref{fig:S7}(d) and Fig.\,\ref{fig:S7}(e).
	\begin{figure}
		\includegraphics[width=1\linewidth]{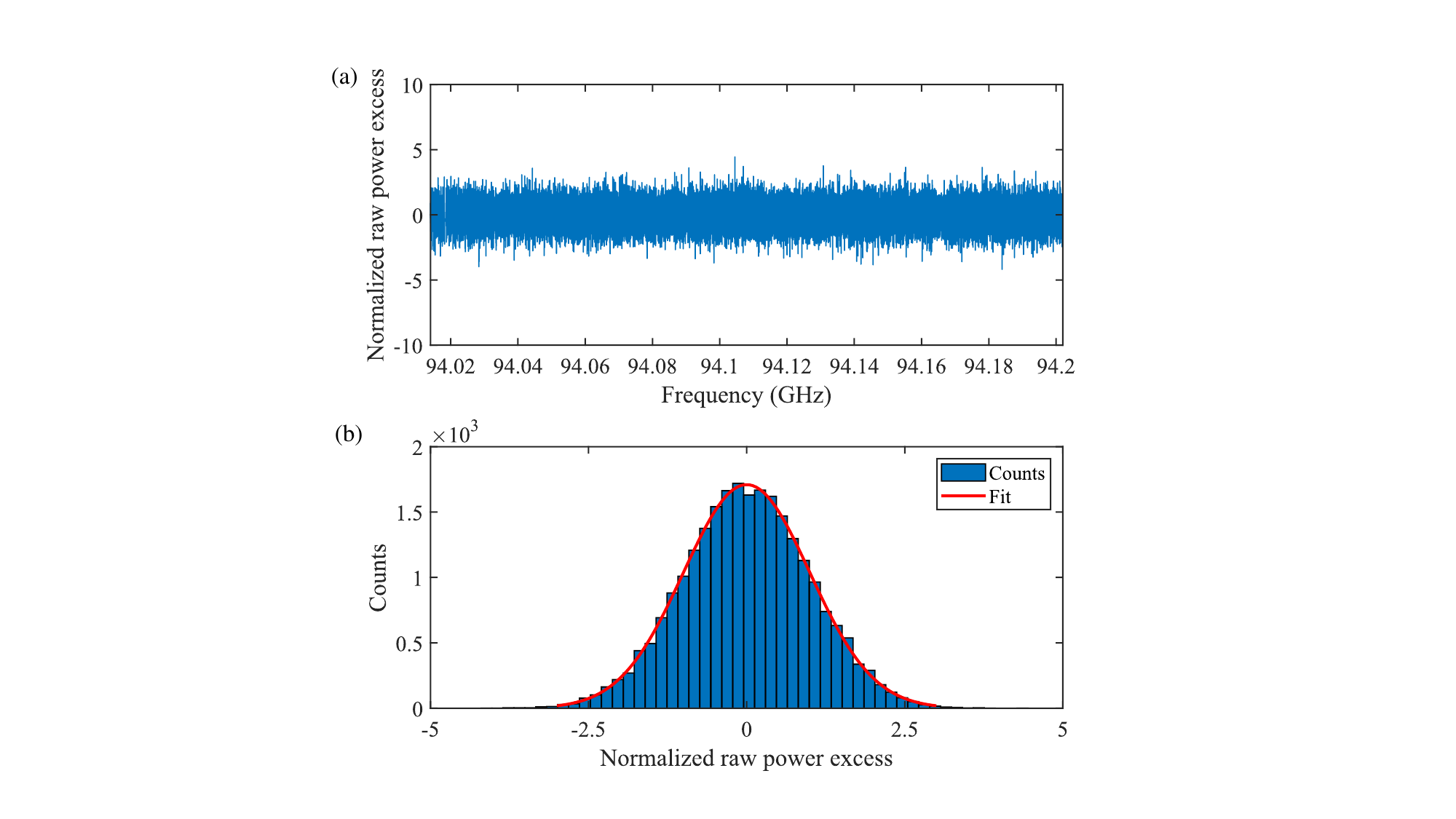}
		\caption{\label{fig:S8} (a) The normalized raw power excess $\Delta_{\rm{norm,raw}}$. (b) The distribution of normalized raw power excess and fitting result with unit Gaussian distribution.}
	\end{figure}
	The bins occupied by artificial signals were subsequently excluded from the raw power excess. The normalized raw power excess $\Delta_{\rm{norm,raw}}$ was obtained by dividing the standard deviation $\sigma_{\rm{raw}}$ of raw power excess. As shown in Fig.\,\ref{fig:S8}(a), there are no bins exceeding $5\sigma_{\rm{raw}}$ in $\Delta_{\rm{norm,raw}}$. The distribution of normalized raw power excess $\Delta_{\rm{norm,raw}}$ fitted well with the unit Gaussian distribution, with mean $\mu=0.000\pm0.012$ and standard deviation $\sigma=1.000\pm0.009$ obtained from fitting result, as illustrated by the rad line in Fig.\,\ref{fig:S8}(b). The raw power excess was multiplied by the averaged noise power $P_n$ to be related to the real power. Then the raw power excess was rescaled to the dark photon signal with $\chi=1$, as shown in Fig.\,\ref{fig:S9}(a). The rescaled power excess $\Delta_{r}$ is represented as
	\begin{equation}
		\Delta_{r} = \Delta_{\rm{raw}}\frac{P_n}{P_s(\chi=1)} ,
		\label{eq:S2}
	\end{equation}
	where $P_s(\chi)$ is the expected dark photon signal power with a specific $\chi$, and the attenuation factors $\eta_{\rm{SG}}$, $\eta_f$ and $F_{\rm{RC}}$ were considered when calculating $P_s(\chi=1)$. To enhance the SNR, the rescaled power excess $\Delta_{r}$ was convolved with the dark photon line shape, which can be described by the Maxwell-Boltzmann distribution:
	\begin{equation}
		\mathcal{F} = 2\sqrt{\frac{f-f_a}{\pi}} \left[ \frac{3}{\eta_c f_a \langle(v_a/c)^2 \rangle} \right]^{3/2} \exp{\left( \frac{-3(f-f_a)}{\eta_c f_a \langle(v_a/c)^2 \rangle}\right)} ,
		\label{eq:S3}
	\end{equation}
	\begin{figure}[h]
		\includegraphics[width=1\linewidth]{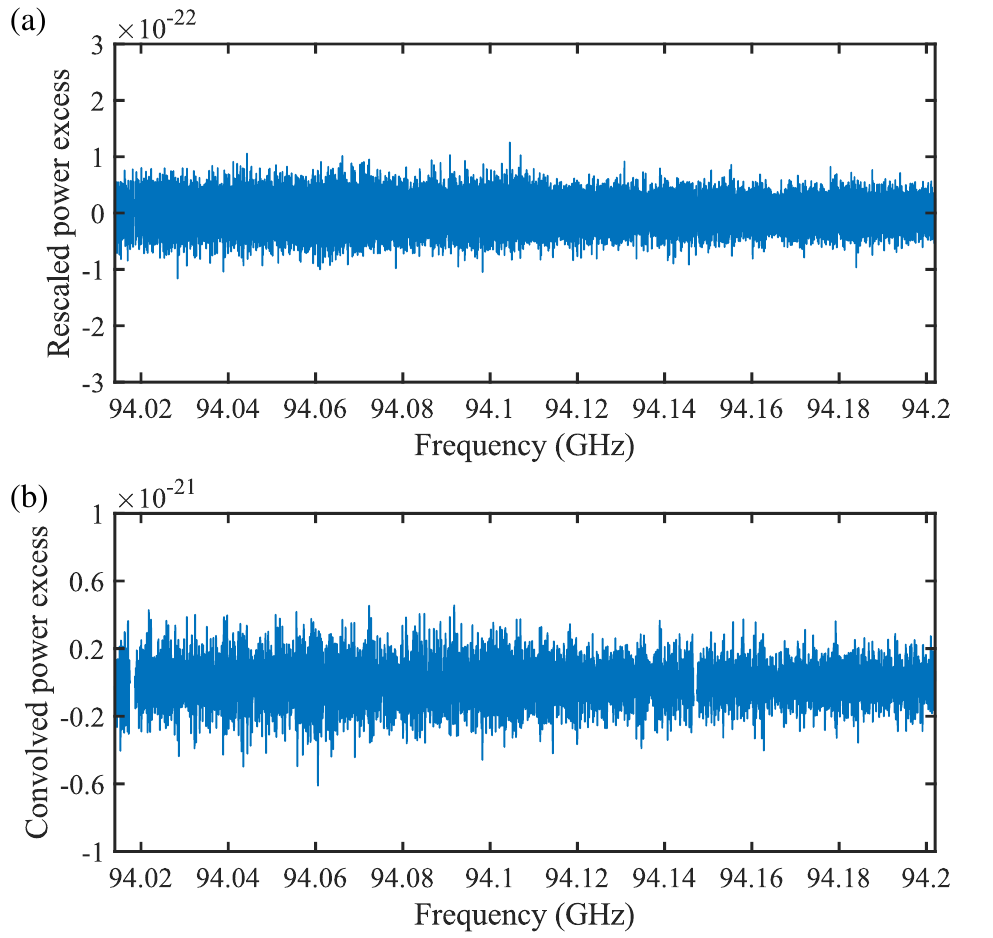}
		\caption{\label{fig:S9} (a) The rescaled power excess $\Delta_{r}$. (b) The convolved power excess $\Delta$. Since the window length of convolution was 100 bins, extra 100 bins near artificial signals were removed during the convolution.}
	\end{figure}
	where $f_a$ is the frequency of dark photons, $c$ is the velocity of light, $v_a$ is the velocity of dark photons, and $\eta_c$ is a correction factor introduced by the complex motion of the lab frame. Here $\langle v_a^2 \rangle = (270\,\rm{km/s})^2$ and $\eta_c=1.7$ was adopted \cite{krauss_calculations_1985,turner_periodic_1990,brubaker_haystac_2017}. The convolution window was set as $763\,\rm{kHz}$ or 100 bins. Then, we obtained the convolved power excess $\Delta$ as shown in Fig.\,\ref{fig:S9}(b). The standard deviation $\sigma_c$ of convolved power excess was also obtained from the convolution on $\sigma_{\rm{raw}}$. The normalized power excess $\Delta_{\rm{norm}}=\Delta/\sigma_c$ is shown in Fig\,\ref{fig:S10}. 
	\begin{figure}[h]
		\includegraphics[width=1\linewidth]{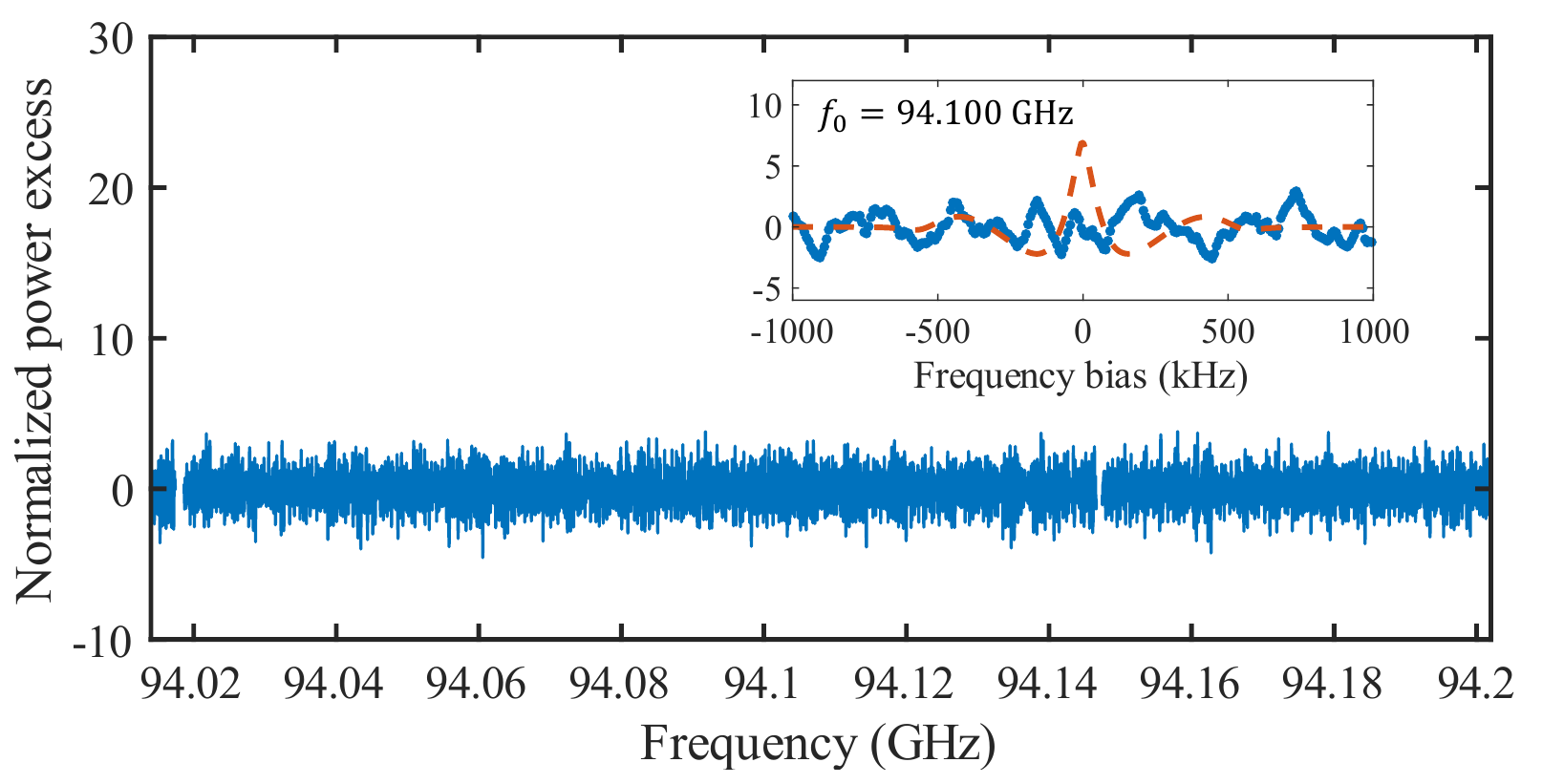}
		\caption{\label{fig:S10} The normalized power excess $\Delta_{\rm{norm}}$ of frequency range 4. The inset shows a hypothetical dark photon signal with $\chi = 3\times 10^{-11}$ at $94.100 \,\rm{GHz}$ (red dashed line), compared with the observation data (blue dots). }
	\end{figure}
	A hypothetical dark photon signal with $\chi = 3\times 10^{-11}$ at $94.100 \,\rm{GHz}$ is shown in the inset of Fig.\,\ref{fig:S10}. The hypothetical dark photon signal was obvious and exceeded $5\sigma$. 
	
	The systematic uncertainties are listed in Table.\,\ref{tab:table1}, and the total systematic uncertainty $\sigma_{\rm{sys}}$ was obtained by adding the individual uncertainties in quadrature. Then the total uncertainties $\sigma'$ were given by
	\begin{equation}
		\sigma'=\sqrt{\sigma_c^2+\Delta^2\sigma_{\rm{sys}}^2} .
		\label{eq:S4}
	\end{equation}
	The total uncertainties $\sigma'$ were utilized to set constraints on kinetic mixing $\chi$.
	
	In this work, Bayesian analysis was employed to provide constraints of the kinetic mixing $\chi$ \cite{cervantes_admx-orpheus_2022} . For a given dark photon signal $P_s$, the measured power excess $\Delta_m$ followed the distribution as
	\begin{equation}
		p(\Delta_m|P_s)=\frac{1}{\sqrt{2\pi}\sigma_m}\exp{\left[-\frac{(\Delta_m-P_s)^2}{2\sigma_m^2}\right]} ,
		\label{eq:S5}
	\end{equation}
	where $\sigma_m$ is the uncertainty of measured power excess $\Delta_m$. Since the dark photon signal $P_s$ was in the unit of $P_s(\chi=1)$, it could be replaced by $\chi^2$:
	\begin{equation}
		p(\Delta_m|\chi^2)=\frac{1}{\sqrt{2\pi}\sigma_m}\exp{\left[-\frac{(\Delta_m-\chi^2)^2}{2\sigma_m^2}\right]} ,
		\label{eq:S6}
	\end{equation}
	By employing the Bayesian analysis theory, the distribution of $\chi^2$ could be represented as \cite{kang_near-quantum-limited_2024}
	\begin{equation}
		p(\chi^2|\Delta)=\frac{p(\Delta|\chi^2)}{\int_{0}^{+\infty}p(\Delta|\chi^2) d\chi^2} ,
		\label{eq:S7}
	\end{equation}
	where $\Delta_m$ and $\sigma_m$ are replaced by experiment results $\Delta$ and $\sigma'$. Finally, by solving the equation
	\begin{equation}
		\int_{0}^{\chi^2_{90\%}}p(\chi^2|\Delta) d\chi^2 = 90\% ,
		\label{eq:S8}
	\end{equation}
	for each bin, the constraints on kinetic mixing $\chi$ with a confidence level of $90\%$ were obtained. Constraint results of eight frequency ranges were combined by selecting the more stringent constraint at each overlapping frequency bin. The combined constraints are shown in Fig.4 of the main text, there are four gaps at $94.377\,\rm{GHz}$,  $94.452\,\rm{GHz}$,  $94.502\,\rm{GHz}$ and  $94.548\,\rm{GHz}$, which are caused by peaks 12, 15, 16 and 20, respectively.
	\begin{table}
		\caption{\label{tab:table1}%
			Summary of the relative systematic uncertainties of experiment parameters. The first and second values (if necessary) denote the minimum and maximum uncertainties within the frequency range from $93.750\,\rm{GHz}$ to $94.550\,\rm{GHz}$.}
		\begin{ruledtabular}
			\begin{tabular}{lc}
				\textrm{Parameter}&
				\textrm{Relative uncertainty}\\
				\colrule
				Coefficient $\beta$ &  $10.9\%$ \\
				Noise of system $P_n$ &  $0.7\% – 5.9\%$ \\
				Area of Disk $A$\footnote{Note a. $A=\frac{\pi D^2}{4}$, where $D$ is the inner diameter of aluminum holder.} &  $1.2\%$ \\
			\end{tabular}
		\end{ruledtabular}
	\end{table}
	
	\section{\label{sec:6} Other production mechanisms of dark photons}
	\begin{figure*}
		\includegraphics[width=1\linewidth]{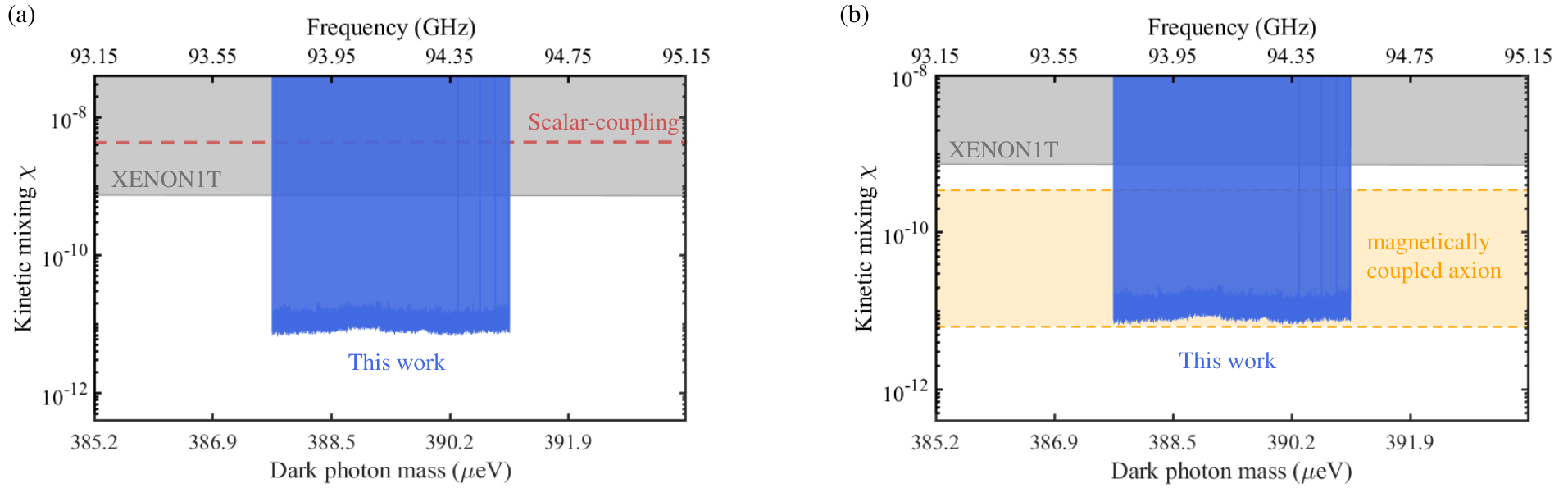}
		\caption{\label{fig:S11} The allowed parameter region of dark photons under different production mechanisms. (a) The red dashed line indicates the upper bound set by the production mechanism through coupling with scalars. (b) The yellow shaded region indicates the allowed parameter region of magnetically-coupled-axion production. The blue shaded region shows the $90\%$ confidence limits set by this work. The gray region refers to the result of XENON1T \cite{aprile_emission_2022}.}
	\end{figure*}
	Dark photons possess varied production mechanisms. Except the quantum-fluctuation production as introduced in the main text, many new production scenarios are proposed very recently. For example, dark photons can be produced through coupling with scalars \cite{cyncynates_detectable_2025}, and there is an upper bound of kinetic mixing due to the cosmology observations, which is $\chi < 4\times10^{-9}$ at $m_{A'}\approx4\times10^{-4}\,\rm{eV}$. As shown in Fig.\,\ref{fig:S11}(a), compared with XENON1T, more parameter region is explored by this work in the mass range from $387.72\,\rm{\mu eV}$ to $391.03\,\rm{\mu eV}$. Additionally, dark photons may also be produced from the magnetically coupled axion \cite{cyncynates_experimental_2025}. In this production scenario, the parameter space of $m_{A'}\approx4\times10^{-4}\,\rm{eV}$, $6.3\times10^{-12}<\chi < 3.5\times10^{-10}$ enables the viable dark photon. As shown in Fig.\,\ref{fig:S11}(b), the allowed region is mostly excluded by this work in the mass range from $387.72\,\rm{\mu eV}$ to $391.03\,\rm{\mu eV}$, and through future enhancement on sensitivity we are able to fully exclude this mechanism in this mass range.
	
\end{document}